\documentclass[%
 reprint,
 amsmath,amssymb,
 aps,
]{revtex4-2}
\usepackage{graphicx}
\usepackage{svg}
\graphicspath{ {./Figures/} }
\usepackage{dcolumn}
\usepackage{bm}
\usepackage{siunitx}
\usepackage[T1]{fontenc}
\usepackage[utf8]{inputenc}
\usepackage{multirow}
\usepackage{mathtools}
\usepackage[shortlabels]{enumitem}

\newcommand{\Tlamp}{T_\mathrm{lamp}}
\newcommand{\QPT}{\mathrm{NUPS}}
\newcommand{\PAT}{\mathrm{PAPS}}
\newcommand{\GammaPAT}{\Gamma_P}
\newcommand{\GammaQPT}{\Gamma_N}
\newcommand{\GammaP}{\Gamma}
\newcommand{\GammaT}{\Gamma}
\newcommand{\nbar}{\bar{n}}
\newcommand{\gother}{g_\mathrm{other}}
\newcommand{\gpat}{g_P} 
\newcommand{\s}{s}
\newcommand{\re}{r}
\newcommand{\zeroflux}{\Phi/\Phi_0 = 0}
\newcommand{\halfflux}{\Phi/\Phi_0 = 0.5}

\newcommand{\peakfluxA}{\Phi/\Phi_0 \approx 0.325}
\newcommand{\peakfluxB}{\Phi/\Phi_0 \approx 0.145}
\newcommand{\xqp}{x_\mathrm{QP}}
\newcommand{\fph}{f_P}
\newcommand{\fq}{f_q}

\newcommand{\Gfrac}{\GammaPAT(\Phi)/\GammaP(\Phi)}
\newcommand{\dD}{\delta \Delta}
\newcommand{\Plamp}{P_{\mathrm{lamp}}}
\newcommand{\mK}{\mathrm{mK}}
\newcommand{\muW}{\mu \mathrm{W}}
\newcommand{\TQPT}{T_\mathrm{ph}}

\newcommand{\g}{|0\rangle}
\newcommand{\e}{|1\rangle}
\newcommand{\Dal}{\Delta_{\mathrm{Al}}}
\newcommand{\Tph}{T_{\mathrm{ph}}}

\newcommand{\pptau}{\langle p(0) p(\tau)\rangle}

\newcommand{\nbarfit}{(1.9 \pm 0.2) \times 10^{-3}}
\newcommand{\dDfit}{4860 \pm 5 \ \mathrm{MHz}}
\newcommand{\fphfit}{112 \pm 2 \ \mathrm{GHz}}
\newcommand{\Gup}{134}
\newcommand{\Gdown}{247}
\newcommand{\patfracmin}{0.53}
\newcommand{\patfracmax}{0.83}
\newcommand{\xqpLfit}{6.2\times 10^{-9}}
\newcommand{\xqpHfit}{0.1 \times 10^{-9}}
\newcommand{\NqpL}{65}
\newcommand{\NqpH}{0.7}

\newcommand{\gratfit}{0.37}
\newcommand{\gotherfit}{8 \ \times 10^{-8} \ \xqp / \mathrm{s}}
\newcommand{\sfit}{11 \ \mathrm{s^{-1}}}
\newcommand{\Tonelim}{2.2}
\newcommand{\minGamma}{81}
\newcommand{\dDfitfull}{4844}

\newcommand{\psdapp}{\mathrm{B}} 
\newcommand{\vardelayapp}{\mathrm{C}}

\newcommand{\modelapp}{\mathrm{E}}
\newcommand{\lampapp}{\mathrm{F}}
\newcommand{\calcgammaapp}{\mathrm{A}}
\newcommand{\singlefapp}{\mathrm{A}}

\newcommand{\burstapp}{\mathrm{H}}
\newcommand{\periodapp}{\mathrm{B}}
\newcommand{\paramsensitivityapp}{\mathrm{G}}
\newcommand{\imageapp}{\mathrm{D}}
\begin{document}

\preprint{APS/123-QED}

\title{Distinguishing parity-switching mechanisms in a superconducting qubit}

\author{S.~Diamond}
\email{spencer.diamond@yale.edu}
\affiliation{Departments of Physics and Applied Physics, Yale University, New Haven, CT 06520, USA}
\author{V.~Fatemi}
\email{Current address: Department of Applied Physics, Cornell University, Ithaca, NY 14853}
\affiliation{Departments of Physics and Applied Physics, Yale University, New Haven, CT 06520, USA}
\author{M.~Hays}
\email{Current address: Department of Electrical Engineering and Computer Science,
Massachusetts Institute of Technology, Cambridge, MA 02139, USA}
\affiliation{Departments of Physics and Applied Physics, Yale University, New Haven, CT 06520, USA}
\author{H.~Nho}
\affiliation{Departments of Physics and Applied Physics, Yale University, New Haven, CT 06520, USA}
\author{P.~D.~Kurilovich}
\affiliation{Departments of Physics and Applied Physics, Yale University, New Haven, CT 06520, USA}
\author{T.~Connolly}
\affiliation{Departments of Physics and Applied Physics, Yale University, New Haven, CT 06520, USA}
\author{V. R.~Joshi}
\affiliation{Departments of Physics and Applied Physics, Yale University, New Haven, CT 06520, USA}
\author{K.~Serniak}
\email{Current address: MIT Lincoln Laboratory, Lexington, MA 02421, USA}
\affiliation{Departments of Physics and Applied Physics, Yale University, New Haven, CT 06520, USA}
\author{L.~Frunzio}
\affiliation{Departments of Physics and Applied Physics, Yale University, New Haven, CT 06520, USA}
\author{L.~I.~Glazman}
\affiliation{Departments of Physics and Applied Physics, Yale University, New Haven, CT 06520, USA}
\author{M.~H.~Devoret}
\email{michel.devoret@yale.edu}
\affiliation{Departments of Physics and Applied Physics, Yale University, New Haven, CT 06520, USA}

\begin{abstract}
Single-charge tunneling is a decoherence mechanism affecting superconducting qubits, yet the origin of excess quasiparticle excitations (QPs) responsible for this tunneling in superconducting devices is not fully understood.
We measure the flux dependence of charge-parity (or simply, ``parity'') switching in an offset-charge-sensitive transmon qubit to identify the contributions of photon-assisted parity switching and QP generation to the overall parity-switching rate.
The parity-switching rate exhibits a qubit-state-dependent peak in the flux dependence, indicating a cold distribution of excess QPs which are predominantly trapped in the low-gap film of the device. 
Moreover, we find that the photon-assisted process contributes significantly to both parity switching and the generation of excess QPs by fitting to a model that self-consistently incorporates photon-assisted parity switching as well as inter-film QP dynamics.
\end{abstract}         

\maketitle
\section{introduction}
A growing sector of electromagnetic radiation sensing and quantum information science rely on superconducting circuits due to their dissipationless nature.
However, nonequilibrium quasiparticle excitations (QPs) notoriously present in the superconductors can cause dissipation and hinder the performance of superconducting devices.
Nonequilibrium QPs can limit the sensitivity of kinetic inductance detectors~\cite{day_broadband_2003}, ``poison'' charge-sensitive devices such as single-Cooper pair transistors~\cite{tuominen_experimental_1992, aumentado_nonequilibrium_2004, ferguson_microsecond_2006},  and cause decoherence in superconducting qubits~\cite{bouchiat_quantum_1998, nakamura_coherent_1999,mannik_effect_2004, guillaume_free_2004, lutchyn_quasiparticle_2005, lutchyn_kinetics_2006,martinis_energy_2009, lenander_measurement_2011, catelani_relaxation_2011,catelani_decoherence_2012, riste_millisecond_2013, wenner_excitation_2013, catelani_parity_2014, pop_coherent_2014, serniak_hot_2018, vool_non-poissonian_2014}.
Hybrid superconductor-semiconductor architectures are likewise susceptible to QP poisoning, which would limit Andreev qubits~\cite{zazunov_andreev_2003,hays_direct_2018,hays_coherent_2021} and proposed Majorana-based qubits~\cite{lutchyn_majorana_2010,rainis_majorana_2012, aasen_milestones_2016}.

\begin{figure}
\includegraphics[width=1.0\columnwidth]{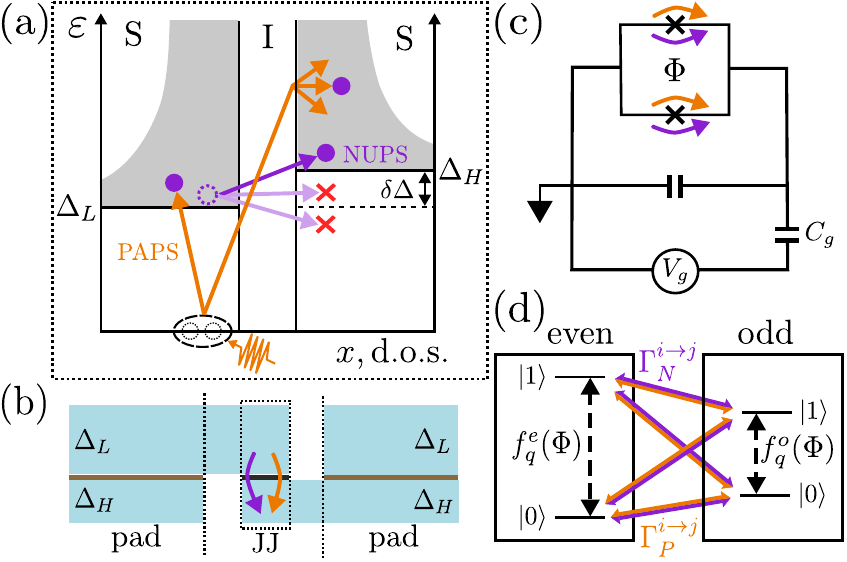}
\caption{ \label{fig:fig1}
(a) Superconducting density of states of the JJ in the excitation picture, with a difference in the superconducting gaps $\delta \Delta$ of the aluminum films. 
Two parity-switching mechanisms are illustrated.
The first conserves QP number while the second generates two QPs. 
Number-conserving parity switching ($\QPT$, purple): a preexisting QP may tunnel while exciting, relaxing, or without exchanging energy with the qubit.
Some  cases of QP-qubit interactions may be suppressed by lack of available final states (red x).
Photon-assisted parity switching ($\PAT$, orange): a photon with energy greater than $\Delta_L + \Delta_H$ may be absorbed at the JJ, breaking a Cooper pair to generate two QPs with one tunneling across the JJ.
(b) Cross-sectional cartoon of the device, formed by one thinner, higher gap film and one thicker, lower gap film. 
Arrows represent $\QPT$ (purple) and $\PAT$ (orange) tunneling across the JJ.
Like the JJ, the pads of the device consist of a bilayer of both films, but have $\sim 10^6 \times$ larger contact area.  
(c)  Circuit diagram of a flux-tunable offset-charge-sensitive transmon. 
Fluctuations in the charge environment may induce offset charge $n_g = C_\mathrm{g} V_\mathrm{g}/2e$.
Both $\QPT$ and $\PAT$ across either JJ result in a switch of the  parity which appears as a sudden jump in $n_g$ by $1/2$.
(d)  Qubit ground ($\left | 0 \right \rangle$) and excited ($\left | 1 \right \rangle$) states separated into even and odd  parity manifolds. 
For $E_\mathrm{J}/E_C \lesssim 30$, $\left| \fq^e - \fq^o \right | \left (n_g = 0 \right ) \gtrsim 500 \ \mathrm{kHz}$.
In addition to causing a parity switch, both $\QPT$ and $\PAT$ may relax or excite the qubit.
}
\end{figure}

QP-induced decoherence in superconducting qubits is typically ascribed to tunneling of excess nonequilibrium QPs across a Josephson junction (JJ), as shown schematically by the purple arrows in Fig.~\ref{fig:fig1}(a).
In this mechanism, when a QP tunnels across the JJ, it couples to the phase across the junction and can thereby cause transitions of the qubit state. 
The rate of QP tunneling occurs in proportion with the QP density in the superconductors, by which previous experiments have inferred QP densities (normalized to the Cooper pair density) in the range $\xqp = n_\mathrm{QP}/n_{\mathrm{CP}} \sim 10^{-9} - 10^{-5}$ ~\cite{palmer_steady-state_2007, shaw_kinetics_2008, lenander_measurement_2011, barends_minimizing_2011, riste_millisecond_2013,pop_coherent_2014,vool_non-poissonian_2014,wang_measurement_2014, gustavsson_suppressing_2016, serniak_hot_2018, serniak_direct_2019, vepsalainen_impact_2020}. 
Depending on levels of other sources of decoherence, QP densities in this range may limit qubit performance.

These QP densities are many orders of magnitude higher than expected for devices at thermal equilibrium with the ${\sim} 30 \ \mK$ base temperature of a dilution refrigerator, in which QP excitations are exponentially suppressed by the superconducting gap $\Delta_{\mathrm{Al}}$.
This contrast points to the nonequilibrium nature of the QPs in the superconductors. 
Furthermore, attempting to explain measurements of QP-induced excitation and relaxation of the qubit by QP tunneling requires an assumption of their presence at high energies~\cite{serniak_hot_2018}.
This is inconsistent with predictions that QPs relax to a distribution near the edge of the superconducting gap~\cite{kaplan_quasiparticle_1976, martinis_energy_2009, serniak_nonequilibrium_2019}.
Despite the many observations of excess QPs across various types of qubits, a two-pronged question remains unanswered: how are nonequilibrium QPs generated and why do they appear to have a nonthermal distribution?

A possible mechanism to help answer these questions was proposed by Houzet et al.~\cite{houzet_photon-assisted_2019}. 
There, it was pointed out that a photon with sufficient energy to break a Cooper pair (${\geq} 2\Delta_\mathrm{Al} \sim 100 \ \mathrm{GHz}$) may be efficiently absorbed at the JJ, generating a pair of QPs. 
Just as QP tunneling results in a single charge being transferred across the JJ, this process likewise switches the parity of the number of electrons that have tunneled across the JJ (referred to simply as the ``parity'' for the remainder of the work) and may cause decoherence of the qubit state [Fig.~\ref{fig:fig1}(a) orange arrows].
We label this process Photon-Assisted Parity Switching ($\PAT$), and it may contribute to QP-induced decoherence in two ways simultaneously: directly, by inducing qubit transitions during parity switches, and indirectly, as a generation mechanism of excess $\xqp$.
These additional QPs may then tunnel and induce decoherence, which we refer to as the NUmber-conserving Parity Switching ($\QPT$) mechanism because unlike $\PAT$, it conserves the number of QPs~\cite{catelani_relaxation_2011, catelani_decoherence_2012}.
We distinguish $\PAT$ from QP generation by ionizing radiation ~\cite{vepsalainen_impact_2020,wilen_correlated_2021,martinis_saving_2021}, which likewise generates QPs but does not directly cause a parity switch. 

Additionally, $\PAT$ causes qubit transitions imitating a nonthermal QP distribution when the typical photon energy is well above $2\Delta_\mathrm{Al}$~\cite{houzet_photon-assisted_2019}. 
For this reason, even in cases where the energy relaxation time of the qubit is limited by other decoherence sources, $\PAT$ may be the primary cause of anomalous excitation of the qubit.
The amount of high frequency radiation that reaches the qubit can be reduced with targeted filtering and shielding, which has been shown to improve qubit performance~\cite{barends_minimizing_2011, serniak_direct_2019, kurter_quasiparticle_2021}. 
Even with these measures in place, recent experiments have shown that stray high frequency photons are indeed absorbed resonantly by spurious antenna modes, inducing parity switching~\cite{rafferty_spurious_2021,pan_engineering_2022,liu_quasiparticle_2022}. 
Given that $\PAT$ can be responsible for QP generation and the appearance of a nonthermal QP distribution, it is imperative to experimentally distinguish the contributions of this mechanism to parity switching and QP generation in superconducting qubits.
However, it is difficult to discern the mechanism from a single measurement of the parity-switching rate or QP-limited energy relaxation time.   

Here, we measure the parity-switching rate $\GammaP$ in a flux ($\Phi$)-tunable transmon sensitive to offset-charge.
As we describe below, the dependence of the parity-switching rate on the applied flux can distinguish $\PAT$ from $\QPT$.
In the flux dependence of the parity-switching rate $\GammaP (\Phi)$, we observed a peak which can be explained by a difference of superconducting gaps between the two aluminum films of the device matching the qubit transition energy.
This gap difference enhances the contrast in flux dependence between $\PAT$ and $\QPT$ and also helps to demonstrate the thermalization of QPs in the device.
Using a new measurement protocol to extract the parity-switching rates conditioned on the initial state of the qubit, we find evidence that excess QPs relax to a low energy distribution and are primarily trapped in the low-gap film.
We developed a model that quantitatively fits the measured $\GammaP (\Phi)$ with a self-consistent combination of $\PAT$ and $\QPT$ and derive two new insights. 
First, $\PAT$ is responsible for a significant fraction of parity switching. 
Second, $\PAT$ generates excess QPs at a rate on par with the sum of all other mechanisms, which we observe by measuring $\GammaP (\Phi)$ in the presence of a controllable photon source operated at several powers.
Due to these effects, estimates of $\xqp$ obtained from measurements of $\GammaP$ or QP-limited qubit relaxation that do not take into account $\PAT$ or gap difference may be inaccurate.
These results advance our understanding of QP dynamics in superconducting qubits and will inform approaches to mitigation of single-charge-tunneling decoherence.

\section{Experimental Device} 

\subsection{Flux-tunable offset-charge-sensitive transmon}
Parity-switching rates were measured directly with an offset-charge-sensitive transmon~\cite{riste_millisecond_2013, serniak_hot_2018,serniak_direct_2019}. 
A parity switching event, in which a single charge tunnels across the JJ, changes the parity $p\in \{e,o\}$ of the electron number in each of its two electrodes. 
Such an event appears as a sudden jump by 1/2 in the reduced
offset-charge $n_g = C_\mathrm{g} V_\mathrm{g}/2e$ (which is measured in units of 2e).
The plasmon eigenstates $i \in \{0, 1\}$ of the qubit can be separated into even and odd  parity manifolds [Fig.~\ref{fig:fig1}(d)], and in the offset-charge-sensitive transmon regime ($E_J/E_C \lesssim 30$), this jump in $n_g$ can result in a substantial jump in the qubit frequency between two values $\fq \pm \delta f_q$, where $\delta \fq(n_g = 0) \gtrsim \mathrm{500 \ kHz}$.
The parity dependence of the qubit frequency has been used previously to measure the parity-switching rates $\GammaP^{i\rightarrow j}$ correlated with qubit transitions from $i$ to $j$, and these rates were inconsistent with parity switching solely due to a thermal distribution of the resident QPs~\cite{serniak_hot_2018}. 

Because the parity-switching rate depends on the density of states available for tunneling [Fig.~\ref{fig:fig1}(a)], changes to the qubit frequency can affect $\PAT$ and $\QPT$ differently.
Thus, we replace the JJ with a DC SQUID in order to tune \emph{in-situ} the average frequency of the qubit $\fq(\Phi)$ with applied flux [Fig.~\ref{fig:fig1}(c)].
For both $\QPT$ and $\PAT$, the flux-dependent rates of parity switching causing a qubit transition from $i$ to $j$ ($\Gamma^{i\rightarrow j}(\Phi)$) depend on the single-charge-tunneling qubit matrix elements as well as factors accounting for the occupation and availability of QP states in the JJ films~\cite{catelani_parity_2014, houzet_photon-assisted_2019, serniak_nonequilibrium_2019}[App. $\calcgammaapp$].
While applying flux changes the single-charge-tunneling matrix elements identically for $\QPT$ and $\PAT$, the resulting change to the qubit frequency $\fq(\Phi)$ affects $\PAT$ and $\QPT$ differently due to the unique constraints on the energies of the QPs involved in each process.     
As a result, the rates of parity switching by $\QPT$ ($\GammaQPT^{i\rightarrow j}(\Phi)$) and by $\PAT$ ($\GammaPAT^{i\rightarrow j}(\Phi)$) can have starkly different dependences on flux.

The offset-charge-sensitive SQUID transmon in this experiment was fabricated with $E_{\mathrm{J}1}/h = 2.465 \ \mathrm{GHz}$, $E_{\mathrm{J}2}/h = 8.045 \ \mathrm{GHz}$.
The large asymmetry of the JJs enabled $\GammaP$ to be measured at all values of flux by mapping the parity onto the state of the qubit~\cite{riste_millisecond_2013,serniak_hot_2018}.
With $E_C/h = 0.352 \ \mathrm{GHz}$, $\delta \fq$ varied from $0.7-14.5 \ \mathrm{MHz}$ as the mean even-odd qubit frequency $\fq $ was tuned from $\fq = 5.0594 \ \mathrm{GHz}$ at $\zeroflux$ to $\fq = 3.5624 \ \mathrm{GHz}$ at $\halfflux$ [App.~$\periodapp$]. 

\subsection{Gap differences in aluminum films}

A key factor influencing the available density of states for parity switching is the difference in the superconducting gaps of the two superconductors on either side of the JJ tunneling barrier. 
The superconducting gap of thin-film aluminum increases with decreasing film thickness ~\cite{chubov_dependence_1969, yamamoto_parity_2006, court_energy_2007}.
Previous works on Cooper pair transistors ~\cite{aumentado_nonequilibrium_2004,court_energy_2007} and Cooper pair box qubits ~\cite{palmer_steady-state_2007, shaw_kinetics_2008} have taken advantage of this effect to trap QPs in thicker aluminum films and reduce QP poisoning of the island.
However, the effect of gap difference on parity switching in transmon qubits has not been previously reported.

The offset-charge-sensitive transmon measured here consists of two aluminum films of 20 nm and 30 nm, respectively~\footnotetext[1]{In bridge-free fabrication of Al/AlOx/Al JJs, the second aluminum film is typically deposited thicker in order to ensure that it climbs the initial film.} [Fig.~\ref{fig:fig1}(a,b), \cite{Note1}].
For these thicknesses, the gaps are expected to differ by $\delta \Delta \coloneqq \Delta_H - \Delta_L \sim 20 \ \mathrm{\mu}e\mathrm{V} \sim  5 \ \mathrm{GHz} \times h $~\cite{chubov_dependence_1969, court_energy_2007}.
This gap difference has two important consequences for parity switching in transmons. 
First, a gap difference changes the proportionality of $\GammaQPT$ to $\xqp$, as certain QP-qubit interactions are suppressed or enhanced depending on the value of $\delta \Delta$ relative to the qubit transition energy $h \fq$ [Fig.~\ref{fig:fig1}(a, purple arrows)]. 
Second, the low-gap film can act as a QP trap, and we show evidence that the excess QPs indeed relax to a cold distribution in the low-gap film [Section III B].

\begin{figure*}
\includegraphics[width=2.0\columnwidth]{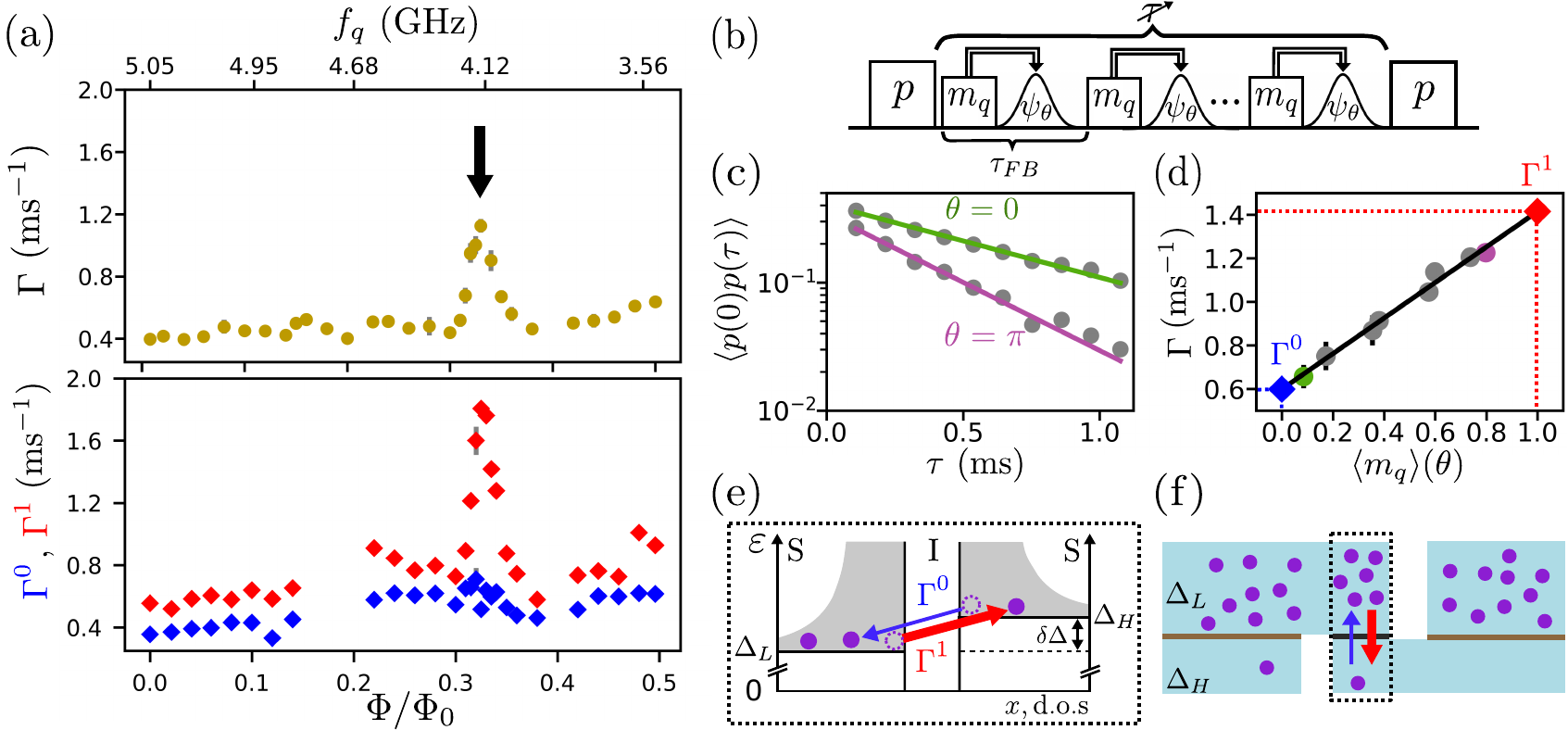} 

\caption{ \label{fig:fig2}
(a) (Upper) Flux dependence of parity-switching rate $\GammaT \approx 0.5 (\GammaT^{0} + \GammaT^{1})$, measured by fitting power spectral densities of jump traces of parity ($\Gamma^{0} \coloneqq \Gamma^{0\rightarrow0} + \Gamma^{0\rightarrow1}, \ \Gamma^{1} \coloneqq \Gamma^{1\rightarrow1} + \Gamma^{1\rightarrow0}$).
Parity is measured by a Ramsey-like sequence which maps the parity onto the state of the qubit~\cite{riste_millisecond_2013, serniak_hot_2018}.
(Lower) Flux dependence of parity-switching rate when the qubit starts in the ground (blue) or excited (red) state, measured by the protocol described in (b).
(b) Pulse sequence for measurement of $\GammaT^{0},\GammaT^{1}$. 
Two measurements of the parity ($p$) are separated by a variable delay of length $\tau$. 
For the duration of the delay, the qubit state is repeatedly measured ($m_q,\ 4 \ \mu \mathrm{s}$), followed by preparation of a superposition  with polarization angle $\theta$ ($\psi(\theta)$), with feedback block repetition time $\tau_{FB} = 5.376 \ \mu \mathrm{s}$. 
(c) The decay of the autocorrelation of the parity $\pptau$ for polarization angles $\theta = 0$ (green) and $\theta = \pi$ (pink) as a function of delay $\tau$ between parity measurements ($\Phi/\Phi_0 = 0.335$).
The parity autocorrelation function $\pptau$ decays at rate $2 \GammaT$, which depends on the $\theta$-dependent time the qubit spends in $|0\rangle$ or $|1\rangle$. 
(d) Measured parity-switching rate $\GammaT(\Phi/\Phi_0=0.335)$ (dots) as a function of the average measurement of the qubit state during $\tau$ $\langle m_q  \rangle$. 
Qubit-state-conditional parity-switching rates $\GammaT^{0}, \GammaT^{1}$ (blue, red diamonds) are determined by extrapolation of a linear fit. 
(e) Schematic depicting $\QPT$ in which the superconducting gap of the films differs by $\dD \approx h f_q$. 
At this flux,  $\GammaQPT^{0\rightarrow1}, \GammaQPT^{1\rightarrow0}$ are strongly enhanced by the divergent densities of states on both sides of the JJ. 
$\GammaT^{0}$ is lower than $\Gamma^{1}$ due to fewer QPs available for tunneling [Fig.~\ref{fig:fig2}(f)].  
(f)  Cross-sectional cartoon of the device as in Fig. 1(b).
Away from the JJ, QPs may tunnel between the two films over a very large contact area.
QPs are shown predominantly in the low-gap film, illustrating the QP trapping effect to which we attribute the rise in $\GammaT^{1}$ but not $\GammaT^{0}$ observed in the lower panel of (a). 
}
\end{figure*}

\section{evidence for gap difference and cold QP energy distribution}

\subsection{Flux-dependent parity-switching rates}
Here we show a signature of gap difference $\delta \Delta$ in the flux-dependent $\Gamma (\Phi)$.
As we explain below, this effect is specific to the $\QPT$ mechanism and aids in the differentiation of $\PAT$ from $\QPT$.
The parity-switching rate was measured as a function of flux through the SQUID loop of the qubit using techniques developed and demonstrated in Refs.~\cite{riste_millisecond_2013, serniak_hot_2018}~[see App. $\psdapp$ for additional detail].
Due to the fast repetition rate and the symmetrized parity-conditional $\pi$-pulses, the qubit spent approximately equal time in the ground and excited states during the experiment.
As a result, the measured parity-switching rate $\GammaP$ weighs the parity-switching rates when the qubit is in $\g$ and $\e$ approximately evenly: $\GammaP \approx 0.5 (\GammaP^{0\rightarrow0} + \GammaP^{0\rightarrow1}) + 0.5(\GammaP^{1\rightarrow1} + \GammaP^{1\rightarrow0})$. 

The flux dependence of $\GammaP$ displays a peak near $\peakfluxA$ ($\fq \approx 4.12 \ \mathrm{GHz}$) on top of a broader increase from $\zeroflux$ -- $0.5$ [Fig.~\ref{fig:fig2}(a, upper)]. 
This broader increase is predicted to result from the flux dependence of the single-charge-tunneling matrix elements.
Here, we focus on the intermediate peak, which may be understood as an enhancement of $\QPT$ when the qubit energy matching the gap difference $h \fq \approx \delta \Delta$.
Under this condition, the rates of $\QPT$-induced relaxation ($\GammaQPT^{1\rightarrow0}$) and excitation ($\GammaQPT^{0\rightarrow1}$) are enhanced due to the divergences of the superconducting density of states on both sides of the JJ, since a QP at the gap edge on one side can tunnel to the gap edge on the other side by exchanging energy with the qubit. 
The implied gap difference $\dD \approx h\fq \approx 20 \ \mu e\mathrm{V}$ is consistent with reported gap measurements for films of these nominal thicknesses~\footnotetext[2]{We note that neither the observed excited state populations nor energy relaxation rates can be responsible for the observed flux dependence of  $\Gamma$.}~\cite{chubov_dependence_1969, court_energy_2007,Note2}. 

This peak indicates that a considerable fraction of parity switching is due to $\QPT$, since $\PAT$ is not enhanced in the same way by $\delta \Delta = h\fq$, as we now explain.
$\PAT$ depends on the sum $\Delta_H + \Delta_L$, rather than the gaps individually, because the sum of the energies of two QPs generated by $\PAT$ are determined by the absorbed photon energy $h\fph$ [Fig.~\ref{fig:fig1}(a), orange arrows].
This is in contrast to $\QPT$, for which the final energy of the tunneling QP is constrained to be the approximately the same as the initial energy, and may only differ by the qubit energy [Fig.~\ref{fig:fig1}(a), purple arrows]. 
The $\PAT$ rate therefore does not exhibit a peak when $hf_q = \delta \Delta$ and shows only a smooth flux dependence due to the matrix elements [App.~$\calcgammaapp$]. 

\subsection{Probing the QP distribution}
The peak in $\GammaP$ also provides insight into the energetic and spatial distributions of QPs in the device.
The average energy of QPs in each film appears to be close to its respective gap, otherwise there would be not be a significant peak when $\delta \Delta = h \fq$. 
This is consistent with predictions that QPs generated at energies above $\Dal$ relax rapidly by emitting phonons toward a steady-state distribution with average energy close to the gap~\cite{kaplan_quasiparticle_1976, martinis_energy_2009, serniak_nonequilibrium_2019}. 

If the QPs are indeed efficiently relaxing to the low energies, it would also be natural to expect that they tend to reside in the low-gap film (i.e., they are trapped there~\cite{riwar_normal-metal_2016, riwar_efficient_2019, pan_engineering_2022}).
The previously described measurement was not directly sensitive to this, so we developed a new experiment to probe the QP densities in each film.
Note that $\GammaQPT^{1\rightarrow0}$ is enhanced by the presence of QPs in the low-gap film and requires the qubit to be in the excited state. 
In contrast, $\GammaQPT^{0\rightarrow1}$ requires QPs in the high-gap film of the JJ and the qubit initially in the ground state.
Thus, by measuring $\GammaP$ with different initial qubit states we can learn about the distribution of QPs in the two films.
\begin{sloppypar}
In our new protocol, the decay of the parity autocorrelation function $\pptau$ was measured with the sequence shown in Fig.~\ref{fig:fig2}(b), which controlled the time the qubit spent in the ground and excited states with active feedback.
During the delay between parity measurements, the qubit undergoes repeated blocks of a qubit state measurement and a preparation into $|\psi_\theta \rangle = \cos{\frac{\theta}{2}} |0\rangle + \sin{\frac{\theta}{2}}|1\rangle$.
In this way, the qubit was projected to the ground (excited) state with probability $\thinmuskip=0mu \cos^2{\frac{\theta}{2}}$ ($\thinmuskip=0mu \sin^2{\frac{\theta}{2}}$) by the ensuing measurement. 
With this procedure, $\pptau \propto e^{-2\GammaP \tau} $ [Fig.~\ref{fig:fig2}(c)], with ${\thinmuskip=0mu~\GammaP~=~\cos^2\frac{\theta}{2}(\GammaT^{0\rightarrow1}+\GammaT^{0\rightarrow0})~+~\sin^2{\frac{\theta}{2}}(\GammaT^{1\rightarrow0}+\GammaT^{1\rightarrow1})}$.
Ideally, polarization angle $\theta = 0$ would keep the qubit in $\g$, and result in measurement of $\Gamma^0\coloneqq \Gamma^{0\rightarrow0} + \Gamma^{0\rightarrow1}$; $\theta = \pi$ would keep the qubit in $\e$ and result in measurement of $\GammaT^{1} \coloneqq \Gamma^{1\rightarrow1} + \Gamma^{1\rightarrow0})$.
In practice, $T_1(\Phi) \approx 20 - 70 \ \mu \mathrm{s}$ limited the experimentally attainable polarizations. 
Instead, the qubit measurement record during the delay was used to estimate the fraction of the delay the qubit spent in each state [App. $\vardelayapp$].
Then, plotting $\GammaP$ as a function of the average qubit state measurement $\langle m_q\rangle$ during the feedback delay, we used a linear fit to extrapolate to $\langle m_q \rangle  = 0, 1$ and infer the parity-switching rate conditioned on the qubit state $|0\rangle$ ($\GammaT^{0}$) and $|1\rangle$ ($\Gamma^1$) [Fig.~\ref{fig:fig2}(d)].
\end{sloppypar}

We repeated this measurement as a function of flux and found that only $\GammaT^{1}$ exhibits a clear peak at $\peakfluxA$ [Fig.~\ref{fig:fig2}(a), lower].
This suggests that the increased parity-switching rate when $h \fq \approx \delta \Delta$ is due to QPs tunneling from the low-gap edge to the high-gap edge by relaxing the qubit [Fig.~\ref{fig:fig2}(e), red]. 
From this, we deduce that generated QPs relax to a relatively cold steady-state distribution with an average energy from the low-gap edge $\langle \varepsilon -\Delta_L \rangle \ll \delta \Delta$.
The thicker, lower gap aluminum film acts as a built-in QP trap [Fig.~\ref{fig:fig2}(f)].

Therefore, the gap difference helps to reduce parity switching in two distinct ways.
In the low-to-high-gap direction, tunneling is reduced because of the lack of available states in the high-gap film.
In the high-to-low-gap direction, tunneling is reduced because the effective $\xqp$ in the high-gap film of the JJ is reduced due to trapping in the low-gap film on top of it. 
For a QP distribution thermalized at a phonon temperature $\Tph \ll \delta \Delta$, the parity-switching rate would be exponentially suppressed while the qubit is in the ground state if $hf_q < \delta \Delta$.
However, in Fig.~\ref{fig:fig2}(a, lower), it is clear that $\GammaT^{0}$ is always of the same order as $\GammaT^{1}$, suggesting that $\PAT$ is also contributing significantly to parity switching in this device.

\begin{figure}
\includegraphics[width=1.0\columnwidth]{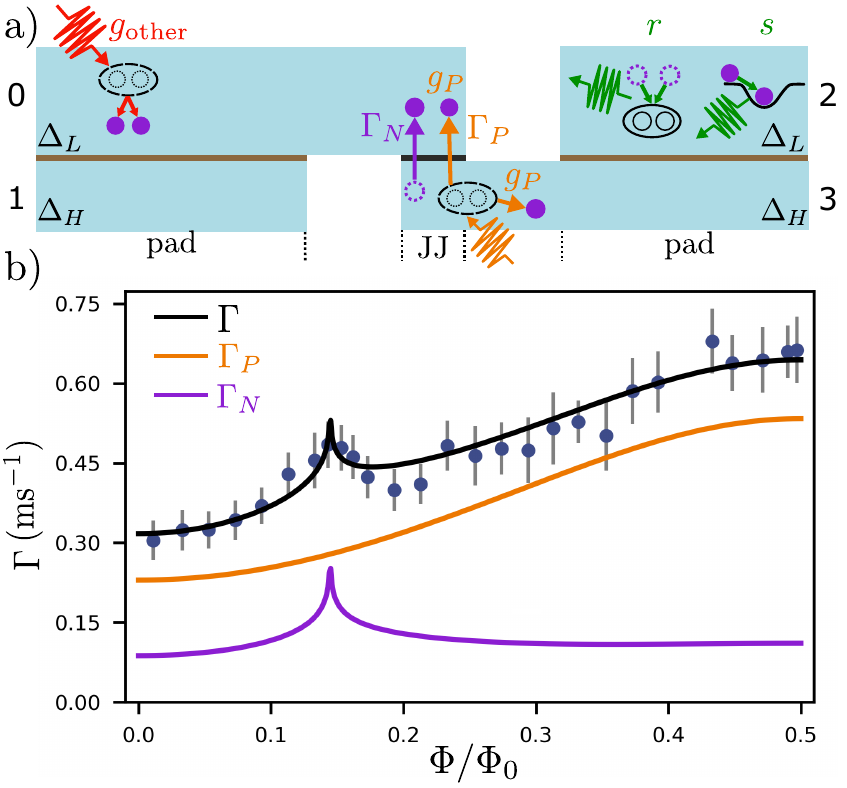} 
\caption{ \label{fig:fig3}
(a) Illustration of QP processes included in our model.
Films are numbered 0-3 for reference in the text. 
Generation of QPs may occur by pair-breaking in the pads (red) at rate $\gother$ or by $\PAT$ at the JJ (orange) at rate $\gpat$. 
QPs tunnel across the JJ at rates $\GammaPAT$ (orange) and $\GammaQPT$ (purple).
QPs are eliminated by recombination at rate $\re$ or are removed from the tunneling population by trapping at rate $\s$ in each film (green). 
(b) Parity-switching rate $\GammaP(\Phi)$ from $\Phi/\Phi_0 = 0 - 0.5$.
The fit to the self-consistent model (Section IV A) is indicated in black, with $\GammaQPT$ (purple) and $\GammaPAT$ (orange) contributions shown explicitly.
The peak near $\peakfluxB$ is explained by the $\delta \Delta \approx h \fq$ condition illustrated in Fig.~\ref{fig:fig2}(e).
}
\end{figure}
\section{PAPS contribution to parity switching}
\subsection{Self-consistent QP dynamics model}
In order to elucidate the contributions of both $\QPT$ and $\PAT$, we developed a model that takes into account the gap difference as well as parity switching and QP generation by $\PAT$. 
With this model, we can take advantage of the different flux dependence of $\QPT$ and $\PAT$ to distinguish between the respective contributions from the fit to $\GammaP(\Phi)$.
The total parity-switching rate is the sum of both parity-switching mechanisms: 
\begin{align}
    \GammaP (\Phi) = \GammaPAT (\Phi, \nbar, \fph) + \GammaQPT (\Phi, \Tph, x_0(\Phi), x_3(\Phi), \delta\Delta). \label{eq:GammaT}
\end{align}
Here, $\nbar$ is the occupation of a high-frequency mode at $\fph$ that couples to the JJ and induces $\PAT$ at a per-photon rate calculated following Ref.~\cite{houzet_photon-assisted_2019}.
In this model, the relative increase of $\Gamma_P$ with $\Phi/\Phi_0$ varying from 0 to 0.5 is determined by $\fph$, while $\nbar$ independently scales the magnitude of $\GammaPAT$.
These are the two fit parameters for the $\PAT$ contribution to $\GammaP$. 
There is an ambiguity in the physical interpretation of these quantities, as it is possible for combinations of modes with varying occupations to give the same effective $\GammaPAT$~[App. $\singlefapp$].
The case for a narrow band of modes with dominant coupling to the JJ has been made previously~\cite{rafferty_spurious_2021,pan_engineering_2022} and is consistent with our data (discussed below).

The $\QPT$ rate depends on the QP densities in the JJ films 0 and 3 ($x_0, x_3$, ``QP'' is dropped for the $\xqp$ of specific films for notation simplicity). 
In an isolated superconducting film, the steady-state $\xqp$ can be determined by balancing QP generation ($g$) with trapping ($s$) and recombination ($r$),
\begin{align}
    \dot{x}_{\mathrm{QP}} &= g - s \xqp - r \xqp^2 = 0 . \label{eq:oldequation}
\end{align}
For our model, we extend this concept to all four films of the device (i.e. the high- and low-gap films on each side of the JJ), considering these dynamics in each film as well as tunneling between films.
We separate generation of QPs into two types. 
Generation by $\PAT$ $g_{P}(\Phi) = \GammaPAT (\Phi) / N_{\mathrm{CP}}, \ $ is included self-consistently with the flux-dependent $\GammaPAT(\Phi)$ contribution to $\Gamma(\Phi)$. 
Other pair-breaking that does not directly result in a parity switch is accounted for by $\gother$, which is assumed to occur equally in each pad.
Trapping at rate $\s$ may arise from vortices or gap inhomogeneities and is included as a fit parameter also assumed to be the same for all four films. 
The recombination rate $r= 1 /(120 \ \mathrm{ns})$ based on the literature~\cite{wang_measurement_2014} is also included in each film.

\begin{sloppypar}
As described earlier [Section III B], the qubit-state dependence of the peak in $\Gamma(\Phi)$ indicates that QPs thermalize and become trapped in the low-gap films  of each pad (films 0 and 2). 
This is supported by predictions that QPs relax by emitting phonons rapidly relative to the other dynamics in the system~\cite{kaplan_quasiparticle_1976, riwar_efficient_2019}.
Under this assumption, we express the QP distributions as Fermi distributions thermalized at $\Tph$, ${f(\varepsilon,\Tph,\mu_{L(R)})=1/(e^{(\varepsilon-\mu_{L(R)})/k_\mathrm{B}\Tph}+1)}$ with nonzero chemical potential $\mu_{L(R)}$ accounting for excess QPs~\cite{palmer_steady-state_2007, glazman_bogoliubov_2021} on the left (right) side of the JJ.  
We take $\Tph \approx 50 \ \mathrm{mK}$ for the temperature of the device based on the reading of a thermometer mounted near the sample cavity [App. $\imageapp$]. 
In the presence of the gap difference, this thermalization will result in the $\xqp$ of the high-gap films being reduced from that of the low-gap films by a factor $e^{-\delta\Delta/k_\mathrm{B}\Tph} \approx 0.008$.
\end{sloppypar}
Because the QPs are assumed to reside predominantly in the low-gap films, we can ignore the effects of trapping and recombination in the high-gap films, which will have a negligible effect on the overall $\xqp$ on each side.
We can thus approximate the dynamics of $\xqp$ in the low-gap films ($x_0$ and $x_2$):
\begin{align}
\dot{x}_{0} &= \gpat + \gother - sx_0 - rx_0^2 - \gamma_{03}x_0 + \gamma_{30}x_2 e^{-\delta \Delta/k_\mathrm{B}\Tph}  \label{eq:x0} \\
\dot{x}_{2} &= \gpat + \gother - sx_2 - rx_2^2 + \gamma_{03}x_0 - \gamma_{30}x_2 e^{-\delta \Delta/k_\mathrm{B}\Tph}  \label{eq:x2}
\end{align}

Here, we have included QP tunneling between films 0 and 3 by the rates $\gamma_{03}$ and $\gamma_{30}$, which are the per-QP tunneling rates in each direction.
These rates take into account $\Tph$ and $\delta \Delta$ and are also flux-dependent.
If the qubit state were in thermal equilibrium with the QPs, $x_0$ and $x_2$ would be the same, because tunneling in each direction across the JJ would be balanced.
However, as discussed in Section III A, the qubit is frequently $\pi$-pulsed throughout the parity-mapping sequence, resulting in nonthermal qubit population during measurement of $\GammaP$. 
The extra time the qubit spends in the excited state results in excess tunneling from film 0 to film 3 as compared to the reverse process, since $\GammaQPT^{10}$ favors tunneling from low-gap to high-gap film due to the densities of states of the superconducting films. 
According to our model, the measurement therefore ``pumps'' QPs from film 0 to 3, which produces a steady-state $\xqp$ in films 2 and 3 that is larger than in same-gap films 0 and 1 \footnotetext[3]{The imbalance of tunneling across the Josephson junction prompts the allowance of $\mu_L \neq \mu_R$}~\cite{Note3}.
This effect is predicted to be particularly strong when $hf_q \approx \delta \Delta$. 

Setting the left-hand side of Eqs.~\eqref{eq:x0} and \eqref{eq:x2} equal to 0 results in coupled equations which can be solved to determine the steady-state $x_0$ and $x_2$.
These values will be flux-dependent due to the flux dependence of $\gpat(\Phi)$ and per-QP tunneling rates $\gamma_{03}(\Phi), \gamma_{30}(\Phi)$.
Finally, $\GammaQPT(\Phi)$ can be calculated given the $\xqp$ of the JJ films ($x_0(\Phi)$ and $x_3(\Phi) = x_2(\Phi)e^{-\delta\Delta/k_\mathrm{B}\Tph}$), and added to $\GammaPAT (\Phi)$ to determine $\Gamma(\Phi)$ [Eq.~\eqref{eq:oldequation}, see App. $\modelapp$ for additional detail on the model]. 

\subsection{Distinguishing parity-switching mechanisms}
In Fig.~\ref{fig:fig3}(b), we show $\GammaP(\Phi)$ for the same device measured during an earlier cooldown, which shows a similar peak as seen in Fig.~\ref{fig:fig2} but at a lower flux corresponding to a value of  $\delta \Delta$ that is  $2.9 \ \mu e \mathrm{V}$ higher.
The cause of this shift is not known, but may be related to mechanisms causing JJ aging~\cite{pop_fabrication_2012}. 
Fitting this data with the self-consistent model (black) yields $\fph = \fphfit$, $\nbar = \nbarfit $, $\delta \Delta / h = \dDfit$.
The values of $\gother$ and $s$ cannot be independently extracted from this fit, so we simply set $\gother=0$ (below this condition will be relaxed, as we will discuss).
The QP densities in the low- and high-gap films that form the JJ (at $\zeroflux$) are $x_0 \approx \xqpLfit$ and $x_3 \approx \xqpHfit$, respectively, corresponding to on average $\NqpL$ QPs in the low-gap film and $\NqpH$ QPs in the high-gap film.
We decompose the fit $\GammaP$ into its $\GammaPAT$ (orange) and $\GammaQPT$  (purple) components and see that the peak at $\peakfluxB$ is due to the effect of $\dD = h\fq$ on $\GammaQPT$, while $\GammaPAT$ is insensitive to $\delta \Delta$ and increases monotonically due to the matrix elements.  
Importantly, we find that $\patfracmin \leq \Gfrac \leq \patfracmax$, indicating that both mechanisms contribute significantly to parity switching. 
\begin{sloppypar}
From the fit to Fig.~\ref{fig:fig3}(b), we extract a QP-induced excitation rate $\GammaP^{0\rightarrow 1}(0) = \Gup \ \mathrm{s}^{-1}$ and QP-induced relaxation rate $\GammaP^{1\rightarrow 0} (0) = \Gdown \ \mathrm{s}^{-1}$.
The ratio ${\GammaP^{0\rightarrow1}/\GammaP^{1\rightarrow0}\approx 0.53}$ implies that QPs cause much more qubit excitation than would be expected by detailed balance at 50 mK, which would predict $\GammaP^{0\rightarrow1}/\GammaP^{1\rightarrow0} \approx 8\times 10^{-3}$.
A previous observation of this type of anomalous excitation was interpreted as evidence of a nonthermal QP distribution of unclear origin~\cite{serniak_hot_2018}.
Here, this apparently nonthermal ratio is interpreted as resulting from $\PAT$. 
\end{sloppypar}
\begin{figure*}
\includegraphics[width=2.0\columnwidth]{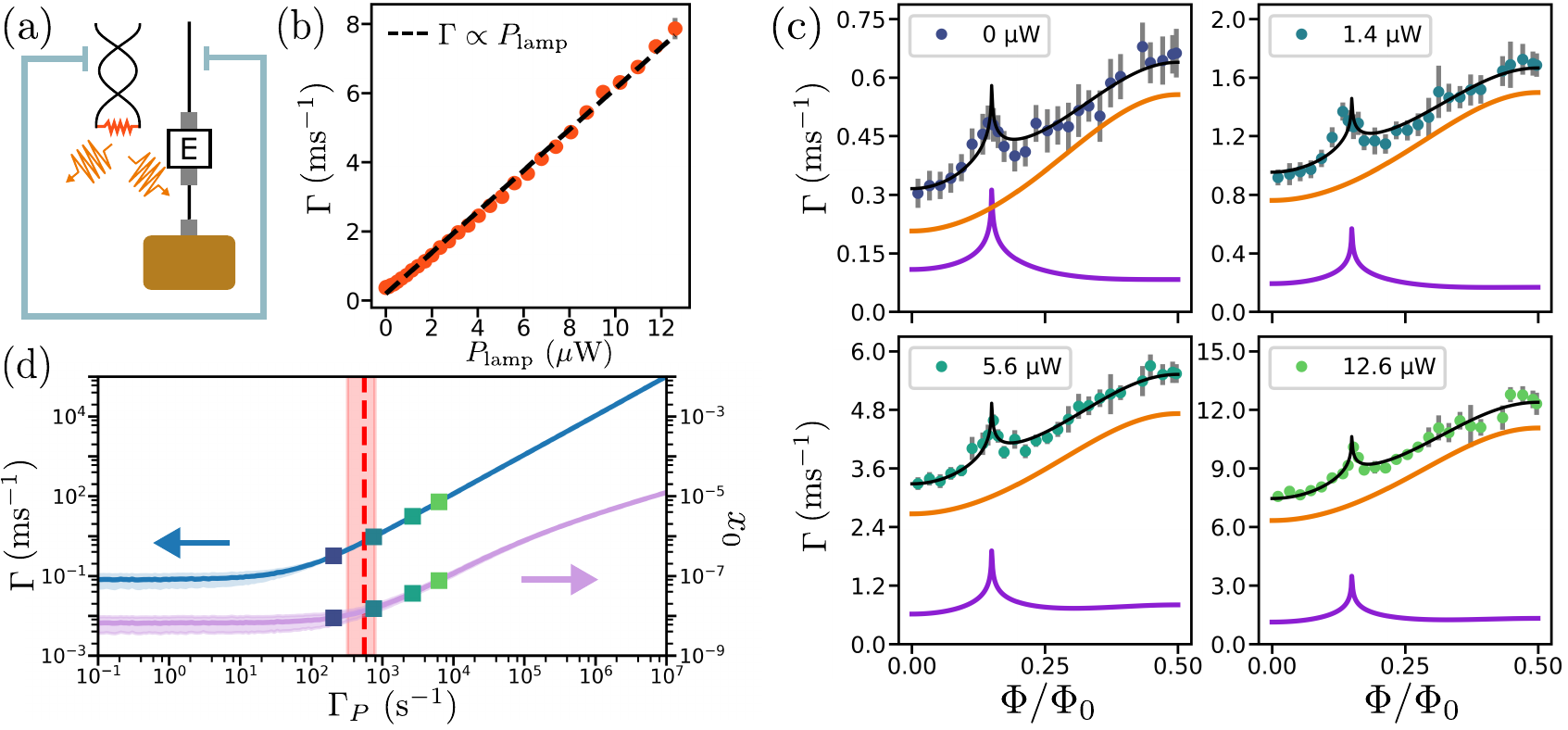} 
\caption{ \label{fig:fig4}
(a) The manganin ``lamp'' resistor hangs in the aluminum can shielding the copper cavity containing the qubit. 
The coax lines include a high-frequency-absorbing Eccosorb CR-110 filter (``E'') in-line just before the readout cavity, inside the can.
Passing current through the lamp causes its temperature to rise, resulting in additional radiation that can leak into the cavity and induce $\PAT$.
Photons may enter the electromagnetic environment of the qubit through SMA connectors or packaging seams.
(b) $\GammaT$ as a function of power dissipated by the manganin ``lamp" resistor, measured at $\zeroflux$.
$\GammaP$ increases approximately linearly with power dissipated by the lamp (black dashed, see App. $\lampapp$ for discussion). 
(c) $\GammaT (\Phi)$ measured at $\Plamp = \{0, 1.4, 5.6, 12.6\} \ \muW$.
We perform a simultaneous fit to all four data sets allowing unique $\fph$, $\nbar$, and shared $\s$, $\gother$, $\delta \Delta$ (solid black).
The contributions of $\PAT$ (orange) and $\QPT$ (purple) indicate that as the lamp power increases, both types of parity switching increase.
(d) Simulated $\GammaT(0)$ (blue, left axis) and $x_0$ (lavender, right axis) as a function of $\GammaPAT$, calculated using fit values from Fig. 4(c).
Experimental values (squares) are marked for reference, and the $\PAT$ level at which $\gpat = \gother$ is marked by the red dashed line. 
Reducing $\GammaPAT$ beyond the current background level ($\Plamp=0)$ would reduce $\Gamma$ and $\xqp$, but the improvement is limited due to $\gother$ which is comparable to $\gpat$.
}
\end{figure*}

\section{QP generation by PAPS}
\subsection{Varying photon incidence}
While the fit in Fig.~\ref{fig:fig3}(b) determines the contribution of $\GammaPAT$ to $\GammaP$, it does not determine whether QPs responsible for the observed $\GammaQPT$ are generated solely by $\GammaPAT$ or if other generation mechanisms also contribute.
The generation contributions $\gother$ cannot be determined from the data shown in Fig.~\ref{fig:fig3} because $\GammaQPT$ depends on $x_0$ and $x_3$, and the trapping rate $s$ can compensate varying levels of $\gother$ to yield the same QP densities.
To obtain an estimate of $\gother$ to compare to $\gpat$, we varied the photon incidence and observed the response in $\xqp$.

To do this, we added a resistor which acted as a controllable source of additional photons for $\PAT$ [Fig.~\ref{fig:fig4}(a)]. 
A $2 \ \mathrm{cm}$ length of manganin wire (${\approx}1.4 \ \Omega$) was suspended inside the aluminum shield containing the copper cavity, though notably not in-line with the coaxial cable for the microwave reflection measurement. 
When current was passed through this ``lamp'' resistor, we observed that $\GammaP (0)$ increased approximately linearly with the power dissipated by the lamp [Fig.~\ref{fig:fig4}(b), App. $\lampapp$]. 
We do not attribute this to an increase in temperature resulting in additional thermal QPs because the temperature measured by a RuOx thermometer installed at the bracket holding the copper 3D readout cavity increased by only a few $\mK$.
Instead, we attribute the increase of $\GammaP$ to an increase in $\PAT$ (including the added generation $g_P$), suggesting that the SMA connections to the cavity and Eccosorb filter inside the aluminum shield may allow photons to leak into the input line~\footnotetext[4]{The same lamp had no effect when implemented outside of the aluminum can surrounding the device}~\cite{Note4}. 

The parity-switching rate $\GammaT (\Phi)$ was measured at $\Plamp = 0 \ \muW$, $1.4 \ \muW$, $5.6 \  \muW$, $12.6 \ \muW$ [Fig.~\ref{fig:fig4}(c)].
Since the overall value of $\GammaP$ increased by $25\times$ from $\Plamp = 0 \ \muW$ to $12.6 \ \muW$ and the intermediate peak at $\peakfluxB$ was still visible, it is clear that $\GammaQPT$ must have increased along with $\GammaPAT$.
This indicates additional $\xqp$ generated by $\PAT$ and constrains $s$ and $\gother$ [App.~$\paramsensitivityapp$]. 
We fit these curves simultaneously with the model described above, allowing unique $\nbar$ and $\fph$ for each $\Plamp$ and assuming common $\s$, $\gother$, and $\delta \Delta$ (fit values shown in Table~\ref{tab:table1}).
The $\fph$ and $\nbar$ values in the table for nonzero lamp power correspond to the additional $\PAT$ added by the lamp on top of the $\Plamp = 0 \ \mathrm{\mu W}$ background.

\begin{table}[b]
\caption{\label{tab:table1}%
Model parameters corresponding to fit shown in Fig.~\ref{fig:fig4}(c).
}
\begin{ruledtabular}
\begin{tabular}{c|ccccc}

$\Plamp$& $\fph$ & $\nbar$ & $\s$ &  $\gother$&$\delta\Delta$ \\
($\muW$) & ($\mathrm{GHz}$) & $\times 10^{-3}$& ($\mathrm{s}^{-1}$ )& $(\xqp /\mathrm{s})\times 10^{-8}$ & ($\mathrm{MHz}$) \\
\hline
$0  $ & $109 \pm \ 3$ & $ \ 2.1\pm .3$ & \multirow{4}{*}{$11 \pm 2$}& \multirow{4}{*}{$8 \pm 3 $} & \multirow{4}{*}{$4844 \pm 3$}\\
$1.4$ & $125 \pm \ 3$ & $\ 2.9 \pm .1$ & \\
$5.6 $ & $125 \pm\ 2$ & $12.8 \pm .4$ & \\
$12.6$ & $124 \pm \ 1$ & $32.6 \pm .8$ & \\
\end{tabular}
\end{ruledtabular}
\end{table}

From this fit, we observe that with the lamp off ($\Plamp = 0 \ \mathrm{\mu W}$), the ratio of generation contributions $\gpat/\gother = \gratfit$, i.e. the rate at which $\PAT$ generates QPs is comparable to generation by other sources.
The extracted trapping rate $s = \sfit$ is of the same order as the trapping rate estimated in~\cite{wang_measurement_2014}. 
Surprisingly, we find that the effective frequency of the additional $\PAT$-generating photons from the lamp is $\fph \approx 125 \ \mathrm{GHz}$ for each power.
If the coupling of high frequency photons to the qubit were broadband, we would expect this frequency to increase as the power of the resistor increased, due to the rising temperature of the emitting blackbody.
Instead, the results may indicate that higher lamp power causes increased occupation of modes that are well-matched to antenna modes of the qubit, which have relatively high absorption efficiency~\cite{rafferty_spurious_2021,pan_engineering_2022, liu_quasiparticle_2022}.
Detailed understanding of the spectrum and coupling of $\PAT$-inducing radiation in 3D transmons are left for future work. 

\subsection{Implications for mitigating parity-switching decoherence}
Having established the QP dynamics, we may extrapolate $\GammaT$ under different levels of $\GammaPAT$-inducing radiation. 
In Fig.~\ref{fig:fig4}(d), we simulate sweeping $\GammaPAT$ for the device measured in this work.  
The blue curve shows the calculated $\GammaT (\Phi = 0)$ with $\gother = \gotherfit$, $\s = \sfit $, and $\delta \Delta = \dDfitfull \ \mathrm{MHz}$ from the fit in Fig.~\ref{fig:fig4}(c), while the lavender curve depicts the corresponding $x_0$ as a function of $\GammaPAT$.
The experimental values derived from the fits in Fig.~\ref{fig:fig4}(c) are marked, and the $\GammaPAT$ at which $\gpat = \gother$ is marked by the red dashed line. 
In the $\gpat \gg \gother$ regime, decreasing $\GammaPAT$ efficiently lowers $\GammaT$. 
High-frequency absorbing filters in the RF lines have indeed been shown to significantly lower $\GammaT$ ~\cite{serniak_direct_2019}. 
The efficient increase of $\GammaP$ with $\Plamp$ emphasizes the previously observed importance of light-tight shielding surrounding the device ~\cite{barends_minimizing_2011, gordon_environmental_2022} to reduce the flux of such photons seen by SMA-connectors below the last in-line filter.

The lower bound due to the QPs generated by non-$\PAT$ mechanisms is $\GammaT \approx \minGamma \ \mathrm{s^{-1}}$, suggesting that removing $\GammaPAT$ entirely would help reduce $\GammaT$ only by a factor of $4$ in this device, with $x_0$ reaching a plateau due to $\gother$. 
To further decrease the parity-switching rate, QPs generated in the pads of the device would need to be addressed.
One such possible source is ionizing radiation, which has been shown to cause correlated errors in qubits across a single substrate attributed to bursts of QPs being generated by the impact and then tunneling~\cite{wilen_correlated_2021}.
In this device, we observe sudden occurrences of rapid parity switching directly [App. $\burstapp$], which supports the interpretation of QPs being the mechanism for these errors. 
The frequency and amount of energy deposited by these bursts as well as the timescale for decay of the generated QP density will determine the extent to which these impacts contribute to $\gother$, which is a subject for future investigation.

An alternative approach is to reduce the harmful impact of QP generation by preventing QPs from tunneling after they have been generated, via QP traps. 
These may be implemented by additional normal-metal or lower-gap superconductor traps~\cite{riwar_normal-metal_2016, riwar_efficient_2019}, but a simpler solution may be to increase the gap difference between the aluminum films.
For fixed number of QPs in the device, the tunneling rate decreases exponentially with the difference between the gaps as discussed above.

The fit to our model yields a limit on energy relaxation time due to QPs of $T_1^{\mathrm{QP}} =\left(\GammaP^{0\rightarrow1} + \GammaP^{1\rightarrow0}\right)^{-1} = \Tonelim \ \mathrm{ms}$ for this device, which is about one order of magnitude longer than $T_1$ of current state-of-the-art transmon qubits~\cite{place_new_2021, gordon_environmental_2022, kurter_quasiparticle_2021}.
While parity-switching decoherence does not currently limit transmons, the eventual reduction of dielectric loss~\cite{martinis_decoherence_2005,siddiqi_engineering_2021} will motivate further mitigation of parity-switching.
Fortunately, parity-switching rates well below the lower limit imposed by non-$\PAT$ sources in this experiment have been measured~\cite{kurter_quasiparticle_2021,gordon_environmental_2022,pan_engineering_2022, iaia_phonon_2022}, with Refs.~\cite{pan_engineering_2022,iaia_phonon_2022} measuring $\Gamma < 1 \ \mathrm{s}^{-1}$
\footnotetext[5]{In Ref.~\cite{pan_engineering_2022}, $\Gamma$ is reduced by engineering of the coupling to high-frequency photons and utilizing larger gap difference between the pads and JJ films as compared to our device. In Ref.~\cite{iaia_phonon_2022}, normal metal reservoirs on the backside of the substrate are shown to aid in reducing $\Gamma$ to lowest-reported levels}\cite{Note5}. 

\section{Conclusions and Outlook}
We have measured the flux dependence of parity switching to distinguish the contributions of Photon-Assisted Parity Switching ($\PAT$) and NUmber-conserving Parity Switching ($\QPT$).
In the flux dependence, we observed a peak which stems from $\QPT$ in the presence of a difference between the superconducting gaps of the aluminum films of our device.
The dependence of this peak in the flux dependence on the qubit state indicates that QPs relax into a low energy distribution in the low-gap aluminum film.
We fit the flux dependence of parity switching with a model that takes into account QP dynamics between the films of the qubit and self-consistent generation of QPs by $\PAT$.
From this fit, we conclude that parity switching in this device is consistent with comparable contributions of $\PAT$ and $\QPT$.
We also found that $\PAT$ generated QPs at a rate similar to other processes that do not directly change the parity.
This work shows that parity switching in transmon qubits cannot be understood in terms of solely $\QPT$ defined by a single QP density. 
The roles of $\PAT$ and gap difference must be considered to accurately determine QP densities from measurements of the parity-switching rate $\Gamma$ or QP-limited energy-relaxation time. 

This device may be modified in several ways in order to better elucidate certain aspects of QP dynamics and generation. 
For example, qubits with different metallization of the pads may result in different spatial distributions of QPs throughout the device. 
Also, changes to the qubit geometry can affect the coupling to $\PAT$-inducing radiation~\cite{rafferty_spurious_2021, pan_engineering_2022, liu_quasiparticle_2022} and aid in the investigation of the spectrum of incident radiation. 
Our self-consistent model, which includes the effects of $\PAT$ and gap difference, can be also be extended to include additional complexities. 
These include possible differences in the superconducting gap at the JJ vs. in the pads~\cite{pan_engineering_2022} or taking into account the full energy dependence of QP dynamics with numerical simulation. 
The framework introduced here will assist investigations of QP generation and impact in these devices going forward.     

\paragraph*{Acknowledgments}
We acknowledge helpful discussions with N. Frattini, A. Koottandavida, G. Catelani, M. Houzet, and R. J. Schoelkopf.
Facilities use was supported by YINQE and the Yale SEAS cleanroom.
We also acknowledge the Yale Quantum Institute.
Research was sponsored by the Army Research Office (ARO), and was
accomplished under Grant Number W911-18-1-0212. 
The views and
conclusions contained in this document are those of the author and
should not be interpreted as representing the official policies, either
expressed or implied, of the ARO or the U.S. Government. 
The U.S. Government is authorized to reproduce and distribute reprints for
Governement purpose notwithstanding any copyright notation herein.
M.H.D. and L.F. are founders and L.F. is a shareholder of Quantum
Circuits Inc.

\paragraph*{Author contributions}
S.D., K.S., M.H., and V.F. designed the device and experimental setup. 
S.D. fabricated the qubit with assistance from L.F.
Author V.R.J. fabricated the parametric amplifier.
S.D. performed the measurements with feedback from V.F., M.H., P.D.K., and T.C.
Authors S.D., V.F., M.H., H.N., P.D.K., T.C., L.I.G., and M.H.D. developed the model and analyzed the data.
S.D., V.F., and M.H.D. wrote the manuscript with feedback from all authors.

\appendix

\section*{appendix a: Calculating $\Gamma_N$ and $\Gamma_P$}
\subsection*{1. Single-charge-tunneling qubit transition rates}
In this section, we derive $\GammaQPT$ and $\GammaPAT$ following Refs.~\cite{catelani_decoherence_2012,houzet_photon-assisted_2019}, and additionally incorporate the flux-dependence of $\Gamma$ for the SQUID device.
The Hamiltonian for single-charge tunneling across the JJ and coupling to the phase degree of freedom $\hat{\varphi}$ of the qubit is
\begin{align} \label{eq:Hcc}
\hat{H}_{\mathrm{QP},\hat{\varphi}} = t\sum_{r,l,s}e^{i\hat{\varphi}/2}\hat{c}_{r,s}^\dagger \hat{c}_{l,s} + \mathrm{H.c.}    ,
\end{align}
where $t$ is the tunneling amplitude, $\hat{c}_{l,s}$ is the electron annihilation operator for the reservoir on the left side of the junction, and $s = \uparrow, \downarrow$ denotes the spin of the electron. 
We apply the Bogoliubov transformation, which diagonalizes the BCS Hamiltonian of the superconductors in the leads.
\begin{align}\hat{c}_{l\uparrow} = u_l\hat{\gamma}_{l\downarrow} +v_l\hat{\gamma}_{l\uparrow}, \ \ \
\hat{c}_{l\downarrow}^\dagger = -v_l\hat{\gamma}_{l\downarrow} +u_l\hat{\gamma}_{l\uparrow}   .
\end{align}
The operator $\hat{\gamma}_{l\downarrow}$ ($\hat{\gamma}^\dagger_{l\downarrow}$) is the operator for annihilation (creation) of a QP excitation on the left with spin down and  $u_l, v_l$ are the conventionally-defined BCS coherence factors, which depend on the QP energy $\varepsilon_l$~\cite{glazman_bogoliubov_2021}.
Applying this transformation to~\eqref{eq:Hcc}, we see that two mechanisms of single-charge tunneling may occur:

\begin{align}
\hat{H}&_{\mathrm{QP},\hat{\varphi}} = t \sum_{r,l,s} \biggl\{  \nonumber \\
&\left[ \left(u_r u_l-v_rv_l\right)\cos{\frac{\hat{\varphi}}{2}}+i\left(u_r u_l+v_rv_l\right)\sin{\frac{\hat{\varphi}}{2}}\right]\hat{\gamma}^\dagger_{rs}\hat{\gamma}_{ls}  \nonumber \\ 
+ &\left[ \left(u_r u_l+v_rv_l\right)\cos{\frac{\hat{\varphi}}{2}}+i\left(u_r u_l-v_rv_l\right)\sin{\frac{\hat{\varphi}}{2}}\right]\hat{\gamma}^\dagger_{rs}\hat{\gamma}^\dagger_{ls} \biggr\}  \nonumber \\
+ &\mathrm{H.c.}   \label{eq:QP_tunneling_H}
\end{align}
The first term, with $\hat{\gamma}^\dagger_{rs}\hat{\gamma}_{ls}$, accounts for $\QPT$ as it empties a QP state on one side and fills a state on the other.
The second term, with $\hat{\gamma}^\dagger_{rs}\hat{\gamma}^\dagger_{ls}$, generates QPs on both sides.
In order to conserve energy, this process may only happen with the absorption of a photon of energy greater than $\Delta_L + \Delta_H$ (the conjugate process annihilates QPs on both sides and emits a photon).
Such radiation will couple to the qubit by imposing a time-dependent phase across the junction. 
Assuming the phase increments $\varphi_P$ induced by the electric field  of the incident photon with frequency $\omega_P$ are small, the single-charge-tunneling operators are transformed by linear expansion of the trigonometric functions in the field-induced phase increments:

\begin{align} \label{eq:trig_ops}
\cos{\frac{\hat{\varphi}}{2}} &\rightarrow \cos{\frac{\hat{\varphi}}{2}} - \varphi_P\sin{\omega_P t} \ \sin{\frac{\hat{\varphi}}{2}} \nonumber \\ 
\sin{\frac{\hat{\varphi}}{2}} &\rightarrow \sin{\frac{\hat{\varphi}}{2}} + \varphi_P \sin{\omega_P t} \ \cos{\frac{\hat{\varphi}}{2}}.
\end{align}

We now apply Fermi's Golden Rule to calculate the rates of parity switching accompanied by a qubit transition from plasmonic eigenstate $i$ to $j$.
Applying Fermi's Golden Rule to the first term of Eq.~\eqref{eq:QP_tunneling_H} gives the rate of $\QPT$ accompanied by such a transition ($\GammaQPT^{i\rightarrow j}$).
The second term of Eq.~\eqref{eq:QP_tunneling_H} gives the rate of $\PAT$ ($\GammaPAT^{i\rightarrow j}$), assuming the presence of high-frequency photons and expanding the single-charge-tunneling operators as above (Eq.~\eqref{eq:trig_ops}).
We also use the Ambagaokar-Baratoff relation to express $t$ in terms of the Josephson energy of the junction $E_J$ and substitute the energy-dependent definitions of $u$ and $v$ above to find:
\begin{align}
\GammaQPT^{i\rightarrow j} = \frac{16 E_J}{\pi \hbar} &\left[ \left|\langle i|\cos{\frac{\hat{\varphi}}{2}}|j \rangle \right|^2 S_{-N} + \left|\langle i|\sin{\frac{\hat{\varphi}}{2}}|j \rangle \right|^2 S_{+N}  \right] \label{eq:GQ} \\
\GammaPAT^{i\rightarrow j} = \frac{\bar{n} g^2 \omega_r}{\pi \omega_q \omega_P} &\left[ \left|\langle i|\cos{\frac{\hat{\varphi}}{2}}|j \rangle \right|^2 S_{-P} + \left|\langle i|\sin{\frac{\hat{\varphi}}{2}}|j \rangle \right|^2 S_{+P}  \right] \label{eq:GP}
\end{align}
Here, $\GammaPAT^{i\rightarrow j}$ is the total $\PAT$ rate induced by photons with frequency $\omega_P$. 
To calculate it, we have multiplied the transition rate induced by a single photon at $\omega_P$ by the average photon number in the mode $\bar{n}$.
The coupling factor which determines the per-photon rate depends on the electric field amplitude and effective dipole length of the qubit. 
Following Ref. \cite{houzet_photon-assisted_2019}, we express this factor in terms of system parameters: geometric coupling rate $g/2\pi = 331 \ \mathrm{MHz}$, readout resonator frequency $\omega_r/2\pi = 9.126 \ \mathrm{GHz}$, and qubit frequency at $\zeroflux$ $\omega_q/2\pi = 5.0594 \ \mathrm{GHz}$.
Here, we only account for $\PAT$ transitions between the ground and first excited states of the qubit.
Photon-assisted transitions to higher states may also occur and add to $\GammaPAT$, but inclusion of these transitions would not qualitatively alter these results. 

The so-called QP structure factors $S_{\pm} = S_\pm^{rl}+S_\pm^{lr}$ include the BCS coherence factors, QP distribution functions, and superconducting density of states $\nu$. 
Tunneling from left to right and right to left are summed to obtain the total rate. 
The structure factors can be expressed:
\begin{align}
S&^{lr}_{\pm N}(\omega_{ij}) = \frac{1}{\bar{\Delta}} \int_0^\infty d\varepsilon \left(1\pm \frac{\Delta_l \Delta_r}{\varepsilon(\varepsilon-\hbar \omega_{ij})}\right) \nonumber \\ 
&\times f(\varepsilon,\TQPT, \mu_l) \nu(\varepsilon, \Delta_l) \nonumber \\
&\times \left(1-f(\varepsilon-\hbar \omega_{ij}, \TQPT, \mu_r)) \nu(\varepsilon-\hbar\omega_{ij}, \Delta_r\right) \label{eq:SQ} \\
S&^{lr}_{\pm P}(\omega_{ij}) = \frac{1}{\bar{\Delta}} \int_0^\infty d\varepsilon \left(1\pm \frac{\Delta_l \Delta_r}{\varepsilon(\hbar\omega_P-\varepsilon-\hbar \omega_{ij})}\right) \nonumber \\
&\times(1-f(\varepsilon,\TQPT, \mu_{l})) \nu(\varepsilon, \Delta_{l}) \nonumber \\
&\times \left(1-f(\hbar \omega_P-\varepsilon-\hbar \omega_{ij},\TQPT, \mu_r)) \nu(\hbar\omega_P-\varepsilon-\hbar \omega_{ij},\Delta_{r}\right) \label{eq:SP}
\end{align}
Nonzero values of $\mu$ in $f(\varepsilon,\Tph,\mu) = \frac{1}{e^{(\varepsilon-\mu)/k_B\Tph}+1}$ describe distributions of QPs which are thermalized at $\Tph$ but have excess number~\cite{palmer_steady-state_2007}. 
The total unassisted and photon-assisted parity-switching rates for a single junction device $\GammaQPT$ and $\GammaPAT$ are then calculated as the sums of the individual rates with qubit transition from $i$ to $j$ weighted by the qubit state probabilities $\rho_0, \rho_1$ 
$\Gamma_m = \rho_0 (\Gamma_m^{0\rightarrow 0}+\Gamma_m^{0\rightarrow1}) + \rho_1 (\Gamma_m^{1\rightarrow 1}+\Gamma_m^{1\rightarrow 0})$,
with $m \in \{N, P\}$. 

In the SQUID transmon, single-charge tunneling across either JJ results in a parity switch.
Therefore, to calculate the parity-switching rates in the SQUID transmon as a function of the flux $\Phi$ through the loop, we sum the respective rates across the JJs with $E_{J1}\ (\Gamma_{m1})$ and  with $E_{J2} \  (\Gamma_{m2})$:
\begin{align} \label{eq:jjsum}
\Gamma_m (\Phi)= \Gamma_{m1}(\Phi)+\Gamma_{m2}(\Phi)
\end{align}
with $m \in \{N,P\}$.
The SQUID transmon Hamiltonian can be expressed in terms of both Josephson energies $\hat{H} = 4E_C(\hat{n}-n_g)^2 - E_{J1}\cos({\hat{\varphi} - \varphi_{\mathrm{ext}}}) - E_{J2}\cos{\hat{\varphi}}$, with the externally tunable flux $\varphi_{\mathrm{ext}} = 2\pi\Phi/\Phi_0$.
Equations \eqref{eq:GQ} and \eqref{eq:GP} are modified to calculate the parity-switching rate for the individual junctions: $ E_J \rightarrow E_{J1},\hat{\varphi} \rightarrow \hat{\varphi} - \varphi_\mathrm{ext}$ for $\Gamma_{m1}$; $\ E_J \rightarrow E_{J2} $ for $\Gamma_{m2}$.
The proportionality factor for photon absorption in Eq.~\eqref{eq:GP} also includes a factor of $\frac{E_{J1(2)}}{E_{J1}+E_{J2}}$, for $\Gamma_{m1(2)}$.
\begin{figure}[h] 
\includegraphics[width=1.0\columnwidth]{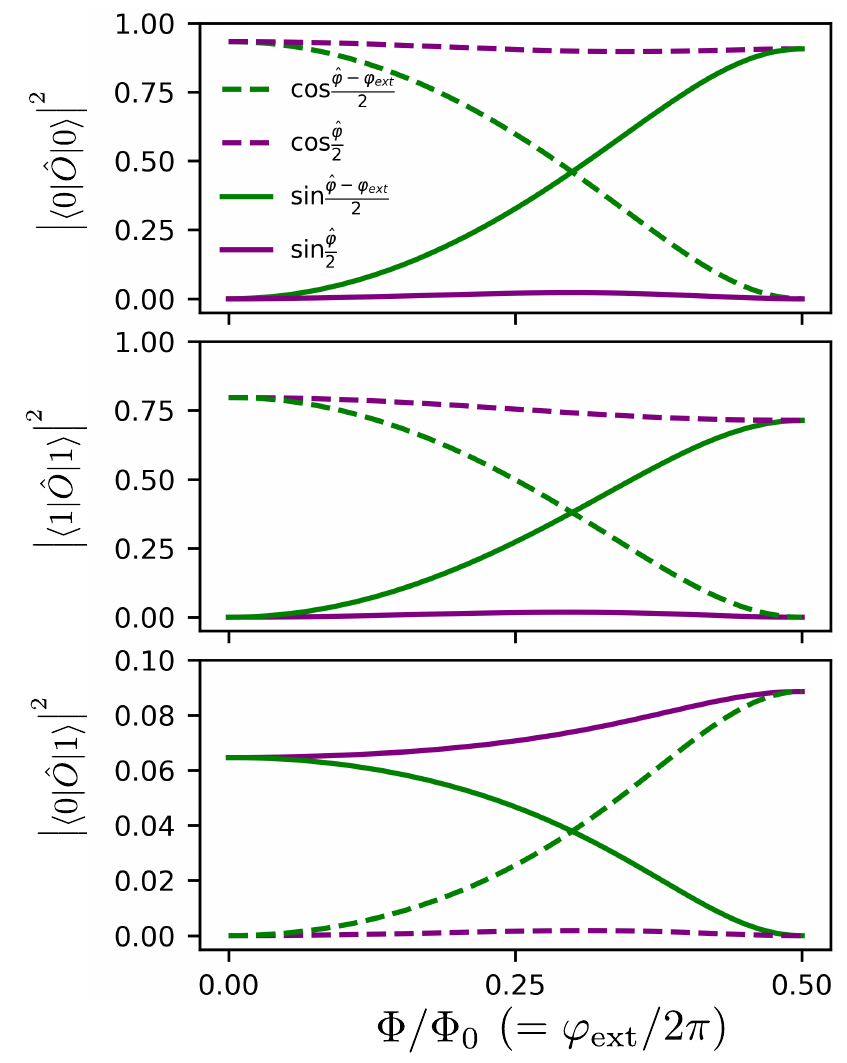} 
\caption{ \label{fig:matelfig}
Qubit matrix elements for single-charge-tunneling as a function of flux through the SQUID loop, with $\Phi/\Phi_0 = \varphi_\mathrm{ext}/2\pi$.
The upper, middle, and lower panels correspond to matrix elements between plasmonic eigenstates $0 \leftrightarrow 0$, $1 \leftrightarrow 1$, and $0 \leftrightarrow 1$.
Green (purple) lines correspond to the matrix elements for single-charge tunneling across the JJ with $E_{J1} \approx 2.5 \ \mathrm{GHz}$ ($E_{J2}\approx 8 \ \mathrm{GHz}$).
}
\end{figure}

The flux dependence of $\Gamma$ results from the tuning of the qubit frequency with flux (see $\omega_{ij}$ in Eqs.~\eqref{eq:SQ}, \eqref{eq:SP}) as well as the flux-dependent matrix elements.
In Fig.~\ref{fig:matelfig}, we plot these matrix elements as a function of flux. 
Green (purple) lines correspond to the matrix elements for single-charge tunneling across the JJ with $E_{J1} \approx 2.5 \ \mathrm{GHz}$ ($E_{J2}\approx 8 \ \mathrm{GHz}$).
The overall increase in $\GammaPAT(\Phi)$ from $\zeroflux$ to $\halfflux$ results primarily from the increase of the $\left| \langle 0 | \sin{\frac{\hat{\varphi}- \varphi_{\mathrm{ext}}}{2}}|0\rangle \right|^2$ and $\left| \langle 1 | \sin{\frac{\hat{\varphi}- \varphi_{\mathrm{ext}}}{2}}|1\rangle \right|^2$ matrix elements for single-charge tunneling across the lower-$E_J$ JJ.  
Note that the wavefunctions $|i\rangle$ implicitly depend on $\varphi_\mathrm{ext}$, ensuring that the choice of assignment for $\varphi_\mathrm{ext}$ does not affect the calculated rates. 

\subsection*{2. Modeling $\GammaPAT$ assuming single photon frequency $\fph$}
\begin{figure}[h]
\includegraphics[width=1.0\columnwidth]{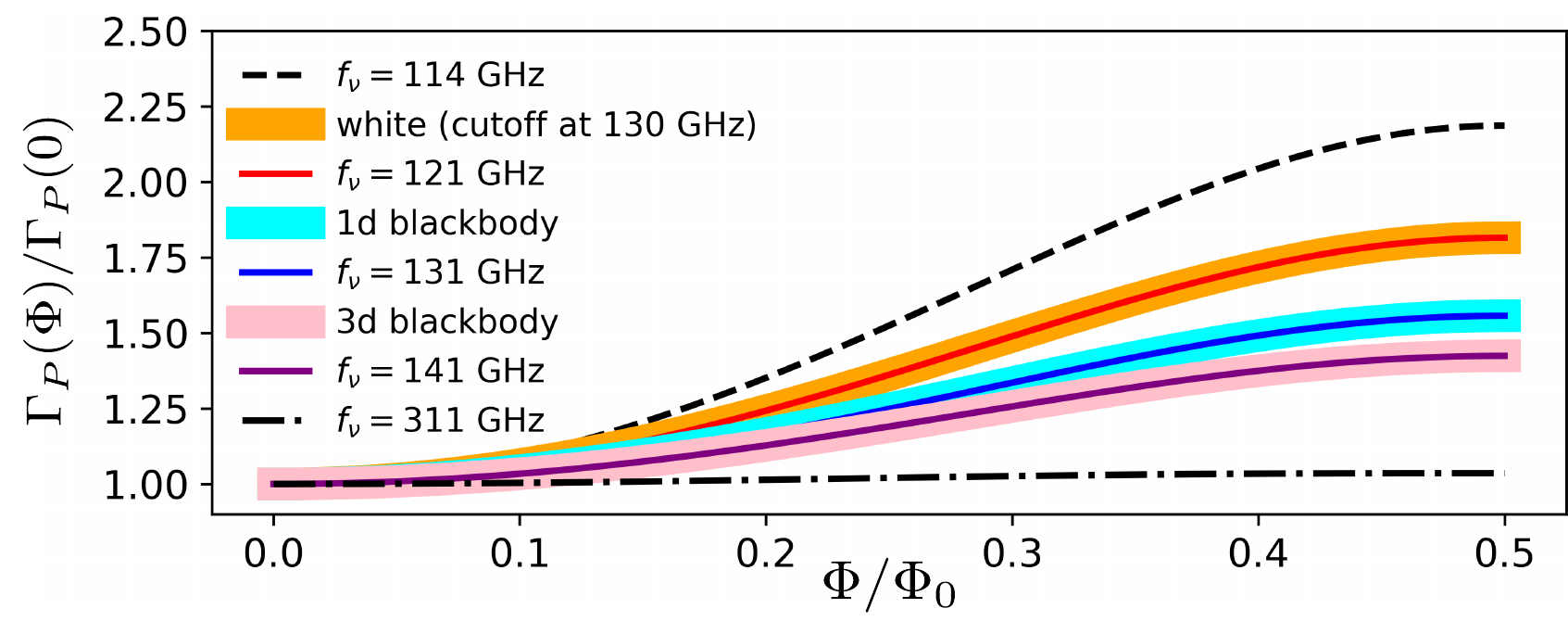} 
\caption{ \label{fig:patfig}
Calculated relative flux-dependence of $\GammaP$ induced by various absorbed photon spectral densities.
Thin lines correspond to $\GammaPAT$ calculated for individual photon frequencies as described in App.~A, with lower photon energies resulting in larger increase in $\GammaPAT(\Phi)/\GammaPAT(0)$.
Thick lines correspond to $\GammaPAT$ calculated for the following spectral densities: white spectrum with cutoff at $2.5 \bar{\Delta} \approx 130 \ \mathrm{GHz}$ (orange), 1d blackbody at 1 K (cyan),  3D blackbody at 1 K (pink). 
Each of these spectral densities result in $\GammaPAT(\Phi)/\GammaPAT(0)$ that can be matched by $\PAT$ induced by individual photon frequencies (thin colored lines).
}
\end{figure}

For the purposes of this work, the true spectrum of the radiation inducing $\PAT$ was not required.
The total induced $\GammaPAT(\Phi)$ for arbitrary spectra of $\PAT$-inducing photons above $\Delta_L +\Delta_H$ can be approximated by an effective occupation $\nbar$ of a single mode at frequency $\fph$: $\GammaPAT(\Phi) =  \GammaPAT(\Phi, \bar{n}, \fph)$. 
Three examples of this are shown in Fig.~\ref{fig:patfig}. 
The relative flux-dependence of the photon-assisted parity-switching rate, $\GammaPAT(\Phi)/\GammaPAT(0)$, depends on the absorbed photon frequency $\fph$. 
Lower $\fph$ induce stronger relative increase with $\Phi$, while $\fph$ cause a weaker relative increase, as shown for 110 GHz (black dashed) and 300 GHz (black dot-dash) in Fig.~\ref{fig:patfig}. 
This is due to smaller photon energies generating QPs closer to the gap, where interference between electron-like and hole-like tunneling of QPs is stronger (see Eq.~\eqref{eq:SP}: lower $\fph$ results in  larger $S_{+P}/S_{-P}$, which emphasizes the flux dependence of the matrix elements). 

In Fig.~\ref{fig:patfig}, we calculate $\GammaPAT(\Phi)$ assuming spectral densities of: a white spectrum with a cutoff frequency of 130~GHz (orange), a 1d blackbody at 1 K from 110 to 300~GHz (cyan) and a 3D blackbody at 1 K from 110 and 300~GHz (pink).
We see that the relative flux-dependence $\GammaPAT(\Phi)/\GammaPAT(0)$ for each spectral density can be fit instead by $\PAT$ induced by photons at a single frequency (solid curves).
The absolute value of $\GammaPAT$ will depend on the unknown attenuation of the radiation as it couples into the cavity, along with the possibly photon-frequency-dependent coupling rate to the qubit. 
Accordingly, for simplicity in our model, $\GammaPAT(\Phi)/\GammaPAT(0)$ is described by a single effective frequency $\fph$, and an average photon number $\nbar$ which determines the magnitude of the $\GammaPAT$ contribution.
Further experiments with different qubit geometries could be performed to elucidate the spectrum and coupling of radiation inducing $\PAT$, as done for 2D qubits in Ref.~\cite{pan_engineering_2022, liu_quasiparticle_2022}.

\section*{appendix b: Measurement of $\Gamma$}
\subsection*{1. Measurement protocol}
\begin{figure}[h] 
\includegraphics[width=1.0\columnwidth]{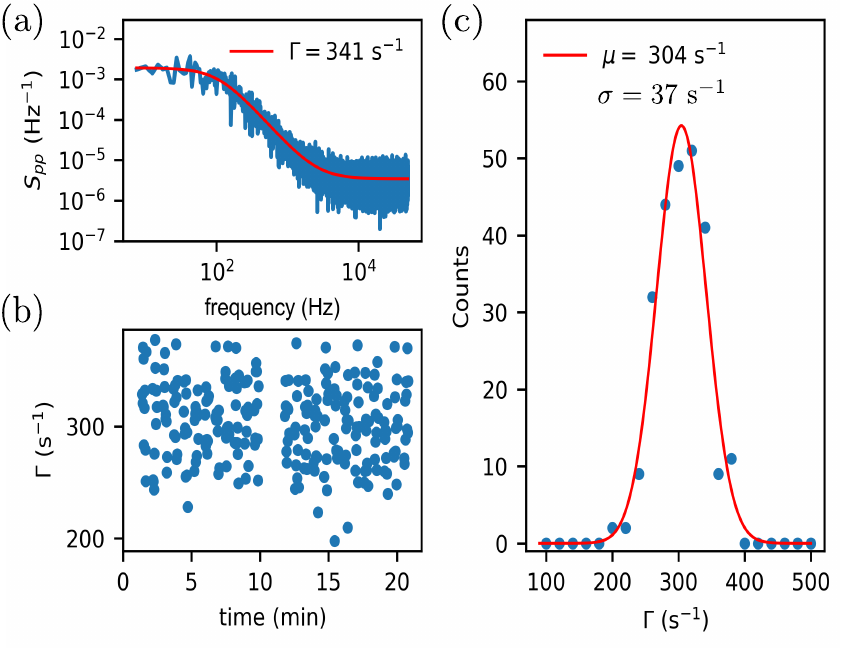} 
\caption{ \label{fig:psd}
(a) Example of a power spectral density of a 400 ms jump trace of parity. 
The power spectral density is fit by a Lorentzian with characteristic rate $\Gamma$, the parity-switching rate, modified by a frequency-independent term due to infidelity of the parity-mapping sequence.
(b) Time series of $\Gamma$ measurements (e.g. Fig.~\ref{fig:psd}(a)), showing the fluctuations of the measured value as a function of time. 
The gap in the data near 11 minutes corresponds to a time when $\Gamma$ was not measurable due to the value of $\delta f$ not meeting the criteria described in the text. 
(c) Histogram of the 250 measurements in (b).
The histogram is fit well by a Gaussian distribution. 
The mean and uncertainty $\mu$ and $\sigma$ give the value and errorbar of the individual $\Gamma$ measurements as plotted in Figs. 2(a, upper), 3(b), 4(b,c).}
\end{figure}

In order to obtain a single measurement of $\GammaP$, we took the power spectral density of a jump trace of parity. 
The jump trace was measured using the Ramsey sequence first demonstrated in Ref.~\cite{riste_millisecond_2013}, which maps the parity onto the qubit state. 
The sign of the second $\pi/2$ pulse in the measurement sequence was alternated between parity measurements such that the sequences enact $\pi$ pulses which alternate being conditioned on the even or odd parity. 
While a measurement of $\GammaP$ could be determined from a jump trace that is ${\approx 2 \ \mathrm {s}}$ long with good signal-to-noise ratio (e.g. Fig.~\ref{fig:psd}(a), in which $\Gamma = 341 \pm 2 \ \mathrm{s^{-1}}$), we observe that these measured values fluctuate in time [Fig.~\ref{fig:psd}(b)].
In order to average over these fluctuations, we measured $\Gamma$ over approximately $20$ minutes.

For the data shown in the main text, we measured 25 jump traces, each comprised of $2 \times 10^6$ measurements of the parity repeated every $10 \ \mu \mathrm{s}$.
Each of these $20 \ \mathrm{s}$ jump traces was chopped into 25 segments that were $400 \ \mathrm{ms}$ long.
We computed the PSD of each segment and averaged five PSDs together to obtain one value of $\GammaP$ [Fig.~\ref{fig:psd}(a)].
Finally, we fit the distribution of all $\GammaP$ measurements with a Gaussian to determine the mean value of $\GammaP$ [Fig.~\ref{fig:psd}(c)], as well as an estimate of the fluctuations from the $\sigma$ of the distribution.
The width of this distribution did not become narrower when more measurements were included, suggesting that the width was due to fluctuation of $\Gamma$ rather than measurement uncertainty. 

In between each jump trace, the difference between the even and odd parity qubit frequencies was checked by a Ramsey experiment.
If the value of $\delta f$ met two criteria, the next jump trace would be measured.
First, the value of $\delta f$ had to exceed a threshold value such that the delay during the parity-mapping sequence $\tau = 1/4 \delta f$ would remain well below $T_2\sim 2 \ \mathrm{\mu s}$.
Measuring at $n_g$ with $\delta f$ below this threshold would result in reduced fidelity of the measurement. 
Second, for fluxes at which the effective  $E_J/E_C \lesssim 20$, the dispersive shifts of the ground and excited states depended on $n_g$~\cite{serniak_direct_2019}. 
In order to measure the parity by mapping onto the qubit state, the ground and excited states needed to have dispersive shifts such that the phase of the reflected measurement signal differed between $\g$ and $\e$.
Therefore, the value of $\delta f$ found by the Ramsey measurement needed to correspond to a value where the ground and excited states were separable. 
This range of usable $\delta f$ values was determined by inspection at each flux prior to measurement of $\Gamma$. 
It was also verified that $\Gamma$ did not itself depend on $\delta f$. 
If the value of $\delta f$ did not meet these two criteria, the dc voltage on the readout pin of the 3D cavity was changed to induce a change in $n_g$ and $\delta f$ was measured again.

\subsection*{2. Measured $\Gamma$ over wider flux range}
\begin{figure}[h]
\includegraphics[width=1.0\columnwidth]{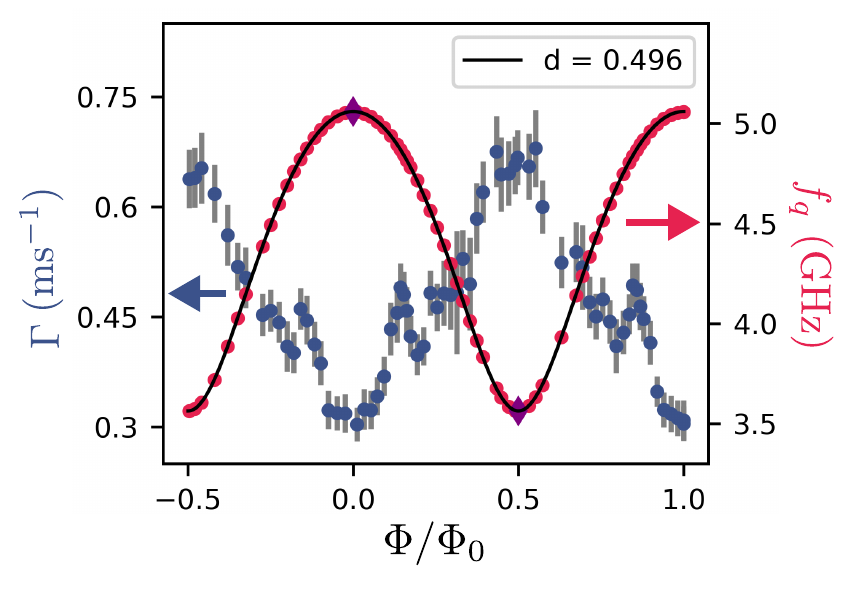} 
\caption{   \label{fig:wideflux}
Charge-parity-switching rate $\Gamma$ (blue, left axis) and mean even-odd parity qubit frequency $f_q$ (pink, right axis) as a function of flux from $\Phi/\Phi_0 = -0.5$ to 1.0. 
The qubit frequencies at $\Phi/\Phi_0 = 0$ and $\Phi/\Phi_0 = 0.5$ were taken to be the maximal and minimal qubit frequencies, respectively, which were measured at the beginning of the cooldown: $f_{q,\mathrm{min}} = 3.5624 \ \mathrm{GHz}$, $f_{q,\mathrm{max}} = 5.0594 \ \mathrm{GHz}$ (purple diamonds). 
The black line shows a fit to just these two data points $f_q(0) = 5.0594 \ \mathrm{GHz}$ and $f_q(0.5) = 3.5624 \ \mathrm{GHz}$ with the expected flux dependence of the frequency $f_q (\Phi) = f_q(0)(\cos^2{\pi \Phi/\Phi_0}+d^2\sin^2{\pi\Phi/\Phi_0})^{1/4}$, where d = $(E_{J1}-E_{J2})/(E_{J1}+E_{J2})$.
The rest of the measurements are assigned flux values based on the qubit frequency according to this model. 
}
\end{figure}

In Fig.~\ref{fig:wideflux}, we show the measured parity-switching rate $\Gamma$ as a function of flux from $\Phi/\Phi_0=-0.5$ to $1.0$ for the cooldown corresponding to the data in Figs.~\ref{fig:fig3}, \ref{fig:fig4}. 
The intermediate peaks corresponding to $h\fq = \delta\Delta$ are also observed at  $\Phi/\Phi_0=-0.145$ and  $\Phi/\Phi_0=0.855$ where this condition is also met. 

\section*{appendix c: measurement of $\Gamma^0, \Gamma^1$}
\begin{figure}[h] 
\includegraphics[width=1.0\columnwidth]{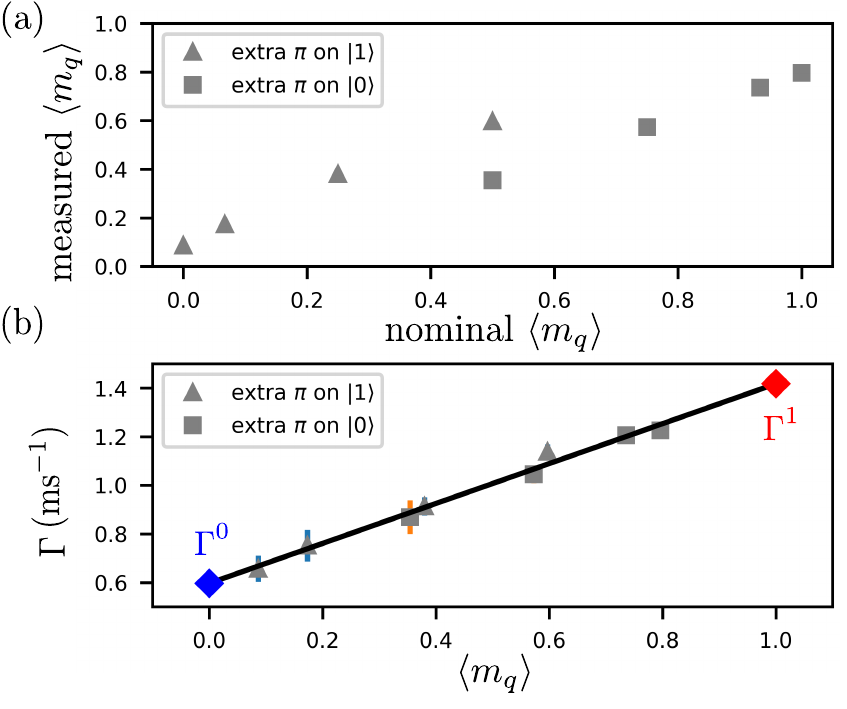} 
\caption{ \label{fig:vardelay}
(a) Average qubit-state measurement during the delay between parity measurements as a function of the nominal polarization angle $\theta$ ($\Phi/\Phi_0 = 0.335$). 
The observed $\langle m_q \rangle$ deviates from the nominal value $\langle m_q \rangle = \sin^2(\theta/2)$  due to energy relaxation as well as the large detuning between the carrier frequency of the qubit pulse and the even and odd parity qubit frequencies.
(b) Charge-parity-switching rate $\Gamma$ as a function of the average qubit-state measurement during feedback between parity measurements [Fig.~2(b)].
For each of the eight measurements of $\Gamma$ (gray), a different state $\psi(\theta)$ is prepared after each qubit measurement $m_q$. 
Triangles represent measurements for which during feedback, the qubit was $\pi$-pulsed if the qubit was in $\e$, followed by rotation to $\psi(\theta)$.
Squares represent measurements for which during feedback, the qubit was $\pi$-pulsed if the qubit was in $\g$, followed by rotation to $\psi(\theta)$.
Both sets of measurements are fit to a single line in order to determine $\Gamma^0$ (blue diamond) and $\Gamma^1$ (red diamond).
}
\end{figure}
In order to measure $\Gamma^{0}$ and $\Gamma^{1}$, the parity-switching rates conditioned on the qubit being in $\g$ and $\e$, we would ideally measure $\Gamma$ with the qubit spending nearly all of the measurement time in $\g$ and $\e$, respectively.
Measurement of $\Gamma$ by fitting the power spectral density of a jump trace of parity does not allow this, because the individual parity measurements in the jump trace necessarily $\pi$-pulse the qubit conditioned on the parity. 
Instead, we measured $\Gamma$ by the decay of the parity autocorrelation function $\langle p(0)p(\tau)\rangle$. 
In this protocol, we separated two measurements of the parity by a variable delay $\tau$. 
During this delay, we repeatedly measured the qubit state and used active feedback to control the mixture of $\Gamma^0$ and $\Gamma^1$ being measured.
The qubit energy relaxation time as a function of flux varied from $20 - 70 \ \mu \mathrm{s}$. 
Since this is only somewhat longer than the time for qubit measurement ($4 \ \mu \mathrm{s}$), the qubit state often changed between repeated measurements.
Therefore, due to these jumps, measuring $\Gamma$ while feeding back to $\g$ or $\e$ still resulted in a mixture of $\Gamma^{0}$ and $\Gamma^{1}$.
While the exact timing of such jumps was unknown, we estimated the fraction of the delay the qubit spent in $\g$ or $\e$ based on the qubit measurement record during $\tau$.
We repeated this measurement for different polarization angles to obtain values of $\Gamma$ with different mixtures of $\Gamma^{0}$ and $\Gamma^{1}$, and we plotted these values of $\Gamma$ vs. $\langle m_q \rangle$, the average qubit measurement during $\tau$.
Then, since $\Gamma \approx \langle m_q \rangle(\Gamma^{1}-\Gamma^{0}) + \Gamma^{0}$, we extrapolated a linear fit of $\Gamma (\langle m_q \rangle)$ to obtain $\Gamma^{0}$, $\Gamma^{1}$.

To summarize, the measurement protocol at each flux point consisted of the following:

\begin{enumerate}[(1)]
\item Perform a Ramsey experiment to determine $\delta f$ for the parity measurements $p$.

\item Measure $p$, delay $\tau$ (during which repeatedly the qubit is measured and prepared into $\psi(\theta)$ every $5.376 \ \mu \mathrm{s}$), measure $p$.
This is done $4 \times$, using all 4 combinations of parity measurements $\{(\pi_e, \pi_e), (\pi_e , \pi_o), (\pi_o, \pi_e), (\pi_o, \pi_o)\}$ where $\pi_e$ and $\pi_o$ are $\pi$-pulses conditioned on the even and odd parity, respectively.

\item Repeat step (2) $150 \times$.

\item Repeat steps (2) and (3) for each $\tau$.

\item Repeat steps (2)-(4) for each $\theta$.

\item Repeat steps (1)-(5) $50 \times$.
\end{enumerate}

Steps (2)-(5) took $\sim 30$ seconds, during which $n_g$ was typically stable. 
If it was found that $n_g$ changed between successive Ramsey measurements, the data between them was not used since the parity measurements were unreliable.
The full protocol took $\sim 1$ hour per flux point.

We now describe how $\theta$ was chosen and implemented in the pulse sequence. 
The extremal cases are simplest.
In order to spend maximal time in $\g$, we prepared $\psi(\theta = 0)$.
The feedback protocol was to $\pi$-pulse the qubit if $m_q = 1$, otherwise do no pulse. 
In order to spend maximal time in $\e$ ($\psi(\pi)$), the protocol was to $\pi$-pulse the qubit if $m_q = 0$, otherwise do no pulse. 
For angles in between, due to the nature of the feedback implementation with the FPGA, only one dynamical angle could be used for rotation.
As a result, the protocol required an additional $\pi$-pulse on either $\g$ or $\e$. 
For example, to prepare $\psi(\pi/4)$, the protocol could be: if $m_q = 0$, rotate by $\pi/4$; if $m_q = 1$, do a $\pi$-pulse and then rotate by $\pi/4$. 
Alternatively, the extra $\pi$-pulse could be enacted instead on $\g$ with $3\pi/4$ rotations to also prepare $\psi(\pi/4)$. 

For the measurement in Fig.~2(d), eight polarization angles were used: $\theta = 0, \pi/6, \pi/3, \pi/2$, with the extra $\pi$-pulse on $\e$ [Fig.~\ref{fig:vardelay}, triangles] and $\theta = \pi/2, 2\pi/3, 5\pi/6, \pi$ with the extra $\pi$-pulse on $\g$ Fig.~\ref{fig:vardelay}, squares]. 
Fig.~\ref{fig:vardelay}(a) shows the average qubit state measurement $\langle m_q \rangle$ for all eight instances of $\theta$ as a function of the expected $\langle m_q\rangle = \sin^2(\theta/2)$. 
We observe that the measured $\langle m_q \rangle$ corresponding to angles in which the extra $\pi$-pulse was acted on $\g$ and those measured with the extra $\pi$-pulse on $\e$ deviate from the expected value. 
We attribute this to two effects. 
First, energy relaxation tends to shift $\langle m_q \rangle$ toward the thermal value. 
Second, large detuning $\delta f$ (up to $\approx 14 \ \mathrm{MHz}$) between the pulse carrier frequency and the even and odd parity qubit frequencies affects the value of $\langle m_q \rangle$ when preparation of $\psi(\theta)$ is performed with the extra $\pi$-pulse. 
For example, $\delta f \approx 5 \ \mathrm{MHz}$ for the data depicted in Fig.~\ref{fig:vardelay}, and thus the qubit state evolves significantly during the time between the $\pi$-pulse and the nominal rotation to $\theta$.
Even with no intentionally added delay between these pulses, there is effectively time between the rotations due to the Gaussian shape of the pulses.
Depending on the specific $\delta f$ and effective evolution time, the true prepared state $\psi$ may differ significantly from the nominally expected value $\psi(\theta)$.

However, calculating the functional dependence of $\langle m_q \rangle$ on $\delta f$ and the pulse lengths was not necessary for this work.
The measurements of $\Gamma^0$ and $\Gamma^1$ relied on measurement of $\Gamma$ with different mixtures of $\g$ and $\e$ and not specific values of $\langle m_q \rangle$.
Fig.~\ref{fig:vardelay}(b) shows that the parity-switching rate $\Gamma$ can be fit by a single line for feedback protocols with the extra $\pi$-pulse on $\g$ or $\e$, since ultimately the value of $\Gamma$ depends on time spent in $\g$ and $\e$ which is estimated by $\langle m_q \rangle$ given the measurement record.
It was also checked that $\psi(0)$ and $\psi(2\pi)$ gave the same values of $\Gamma^0, \Gamma^1$ within measurement uncertainty, showing that the drive power itself does not influence the measured value of $\Gamma$. 

\section*{appendix d: measurement setup and device images}
\subsection*{1. Device images}
The offset-charge-sensitive SQUID transmon is comprised of e-beam evaporated aluminum on a sapphire substrate. 
The fabrication process was the same as that described in detail in the Supplemental Material of Refs.~\cite{serniak_hot_2018,serniak_direct_2019}.

\begin{figure}[h]  \label{fig:images}
\includegraphics[width=1.0\columnwidth]{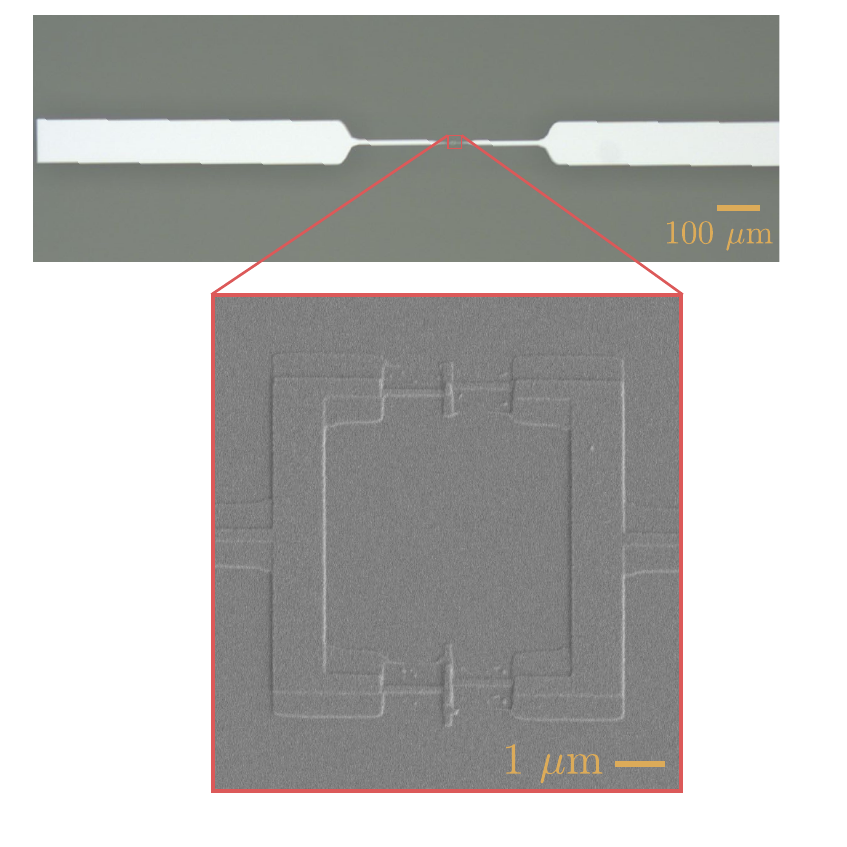} 
\caption{
(upper) Optical image of qubit fabricated on the same wafer as the experimental device, which is nominally identical.
The end of the right pad is not in view, but is symmetric with the pad on the left. 
(lower) Scanning-electron-microscope image of the SQUID loop of the optically imaged device, in which the difference of the areas of the two Josephson junctions are apparent. 
The qubit is fabricated using the bridge-free technique in a liftoff process as described in Refs.~\cite{serniak_hot_2018, serniak_direct_2019}.
}
\end{figure}
\subsection*{2. Measurement setup}

\begin{figure}[h]
\includegraphics[width=1.0\columnwidth]{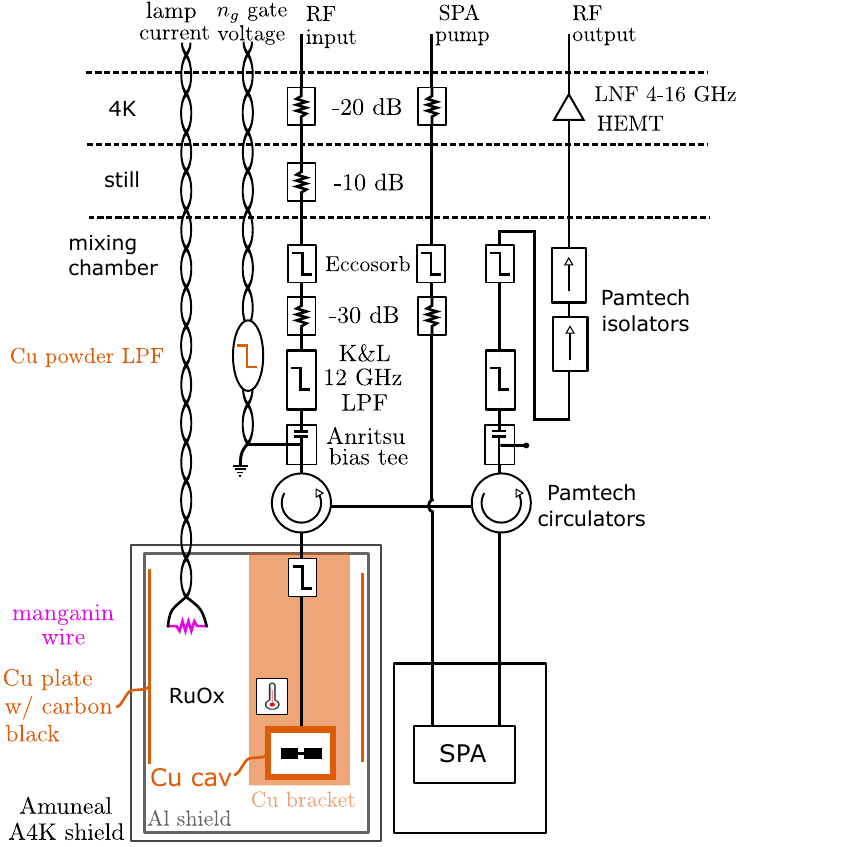} 
\caption{ \label{fig:setup}
Cryogenic measurement setup. 
}
\end{figure}

The qubit was mounted in a 3D Cu microwave readout cavity and measured in reflection ($\kappa = 3.5 \ \mathrm{MHz}$) as diagrammed in Fig.~\ref{fig:setup} with a SNAIL parametric amplifier providing initial amplification to achieve single-shot qubit-state readout. 
An Eccosorb filter inside the aluminum and magnetic shields was included, which has been demonstrated to reduce the parity-switching rate \cite{serniak_direct_2019}. 
A $2 \ \mathrm{cm}$ section of manganin wire ($\approx 1.4 \ \Omega$) was suspended in the aluminum can by superconducting leads and acted as an adjustable source of $\PAT$-inducing photons. 
A RuOx thermometer was mounted on the bracket holding the Cu readout cavity in order to monitor the temperature of the bracket as power was dissipated by the manganin wire. 
A dc voltage bias was added to the RF input line with a pair of bias tees, such that a dc voltage could be applied to the readout pin in the cavity in order to bias the offset charge. 
All lines entered the shields via a narrow slot opening and copper tape was used to make the slot as light-tight as possible.  

\section*{appendix e: four-film model for QP dynamics}
Here, we provide additional detail to Section IV A on the calculation of $x_0$ and $x_3$, the QP densitites of the JJ films that determine $\GammaQPT$. 
We consider the QP dynamics in all four films of the device in order to determine these values, taking into account generation of QPs by $\PAT$ and tunneling across the JJ self-consistently with $\GammaPAT$ and $\GammaQPT$, respectively. 
We first consider $x_0$ and $x_1$, the QP densities of the films on the left side of the JJ in Fig. 3(a). 
These QP densities may change by mechanisms of Eq.~\eqref{eq:oldequation} (generation, trapping, and recombination), but also by tunneling between films of the device.
QPs may tunnel between the low- and high-gap films 0 and 1 ($t_{01}, t_{10}$) as well as across the JJ between films 0 and 3 ($\gamma_{03}x_0, \gamma_{30}x_3$).  
The interfilm transport can be described by the coupled equations:
 \begin{align}
    \dot{x}_0 &= g_0 - sx_0 - rx_0^2 - (t_{01} - t_{10}) - \gamma_{03}x_0 + \gamma_{30} x_3 \label{eq:x0s} \\
    \dot{x}_1 &= g_1 - sx_1 - rx_1^2 + (t_{01} - t_{10}) \label{eq:x1s}
 \end{align}
Below, we describe our approach to these inter-film tunneling terms.  

While films 0 and 1 are separated by an $\mathrm{AlO_x}$ layer, they share a large contact area in the pads, so QPs tunnel rapidly between the films. 
In Section III B, we showed evidence that QPs thermalize to distributions primarily located in the low-gap films on each side of the JJ. 
We therefore assume Fermi distributions for the QPs, $f(\varepsilon, \Tph,\mu_{L})=1/(e^{(\varepsilon-\mu_{L})/k_\mathrm{B}\Tph}+1)$ with films 0 and 1 sharing the same nonzero chemical potential $\mu_{L}$ to describe an excess QP number independent of the temperature $\Tph$~\cite{palmer_steady-state_2007,glazman_bogoliubov_2021}.
The QP densities in each film $x_0$ ($x_1$) are related to $\mu_L$ through the definition of $\xqp = 2 \int_0^\infty d\varepsilon \ \nu(\varepsilon, \Delta_{0(1)})f(\varepsilon, \Tph, \mu_L)$.
Due to the difference between the gaps, this thermalization results in $x_1 \approx x_0 e^{-\eta}$, where $\eta = \delta \Delta / k_\mathrm{B} \Tph$. 
In our model, we therefore replace the tunneling rates between films 0 and 1 ($t_{01}$ and $t_{10}$) with this assumption of rapid thermalization.
We add Eqs.~\eqref{eq:x0s} and \eqref{eq:x1s} and substitute $x_1 = x_0e^{-\eta}$.
The same approach is applied to films 2 and 3, which likewise share a large contact area in the opposite pad, reducing the number of independent QP densities in the system from four to two.
This results in two coupled equations for $x_0$ and $x_2$:
\begin{align}
    \dot{x}_{0} = &\frac{1}{1+e^{-\eta}}[(g_0 + g_1) - (1+e^{-\eta})sx_0\nonumber \\
    &- (1+e^{-2\eta})rx_0^2 - (\gamma_{03}x_0 - \gamma_{30}x_2e^{-\eta}) ] \label{eq:x03} \\
    \dot{x}_{2} = &\frac{1}{1+e^{-\eta}}[(g_2 + g_3) - (1+e^{-\eta})sx_2\nonumber \\
    &- (1+e^{-2\eta})rx_2^2 + (\gamma_{03} x_0 - \gamma_{30} x_2e^{-\eta}) ]     \label{eq:x30}
\end{align}

As described in Section IV A, if the qubit were in thermal equilibrium with the QPs, the films on opposite sides of the JJ would share the same chemical potential. 
However, the qubit state is frequently pulsed during measurement of $\Gamma$, and as a result the tunneling from films 0 to 3 and films 3 to 0 are not equal.
The QP densities on the opposite side of the JJ ($x_0$ and $x_2$) differ in the steady-state during the measurement due to this imbalance.

The rates $\gamma_{03}$ and $\gamma_{30}$ are the per-QP tunneling rates in the 0 to 3 and 3 to 0 direction, respectively.
These may be calculated using only the $S^{lr}$ or $S^{rl}$ terms of $S_{\pm N}$ [Eq.~\eqref{eq:SQ}], and assuming  nonthermal qubit-state weights $\rho_0 = \rho_1 = 0.5$ caused by the measurement sequence. 
Because $\Gamma_{N,03}$ is approximately proportional to $x_0$ and independent of $x_3$ for small $x_3$ such as those existing in this experiment, the per-QP tunneling rate $\gamma_{03} \coloneqq \Gamma_{N,03} / (x_0 N_{\mathrm{CP,0}})$ can be calculated.
Similarly, $\gamma_{30}~\coloneqq~\Gamma_{N,30}/(x_3 N_{\mathrm{CP,2}})$. 
The number of Cooper pairs in each film $N_{\mathrm{CP},k} = 2D(\varepsilon_F)\Delta_k V_k$, with film number $k \in \{0,1,2,3\}$ depends on the superconducting gap and volume. 
The single spin density of states at the Fermi energy $D(\varepsilon_F) = 0.72\times 10^{29} \mathrm{\mu m}^{-3} \mathrm{J}^{-1}$~\cite{court_quantitative_2008} and the volumes are approximately $100 \times 700\times 0.03 \ \mathrm{\mu m}^3$ for the low-gap films and $100 \times 700 \times 0.02 \ \mathrm{\mu m}^3$ for the high-gap films. 

These per-QP tunneling rates depend on $\Tph$ and $\delta \Delta$, and have significant dependence on flux.
The imbalance $\Gamma_{N,03}/\Gamma_{N,30} > 1$ is largest when $h\fq = \delta \Delta$, since QPs at the low-gap edge in film 0 can absorb an excitation from the qubit to tunnel to the high-gap edge in film 3. 
Based on our fit, this results in 55\% more QPs in films 2 and 3 as compared to films 0 and 1 at the flux for which $h\fq = \delta \Delta$.

We separate QP generation into generation by $\PAT$ ($g_P$) and generation by other mechanisms ($\gother$), which break Cooper pairs but do not simultaneously result in a parity switch. 
While $\PAT$ generates QPs only in JJ films 0 and 3, mechanisms contributing to $\gother$ could generate QPs in all the films. 
Therefore, we have $g_0 + g_1 = (g_P + g_{\mathrm{other},0}) + g_{\mathrm{other},1} = g_P + \gother$, where we have defined $\gother \coloneqq g_{\mathrm{other},0} + g_{\mathrm{other},1}$ as the total generation on one side of the JJ.  
Likewise, $g_2+g_3 = g_{\mathrm{other},2} + (g_P + g_{\mathrm{other},3}) = g_P + \gother$, where we have made the reasonable assumption that non-$\PAT$ pair-breaking occurs identically on opposite sides of the JJ since the two pads have the same geometries.  
Generation by $\PAT$ $g_P \coloneqq \GammaPAT/N_{\mathrm{CP},0}$ is calculated self-consistently with $\GammaPAT$ and $\gother$ is a flux-independent fit parameter in our model. 

In Section IV A, the additional approximation is made that since $e^{-\eta}\approx 0.008$ for $\Tph \approx 50 \ \mathrm{mK}$ and $\delta \Delta \approx 4.844 \ \mathrm{GHz} \approx 233 \ \mathrm{mK}$, the high-gap films can essentially be removed from the determination of $x_0$ and $x_2$. 
Eqs.~\eqref{eq:x0} and \eqref{eq:x2} are obtained by substituting the above expressions for generation  into~\eqref{eq:x03} and \eqref{eq:x30}, and making the approximation $e^{-\eta} = 0$. 

\section*{appendix f: effect of lamp}
\subsection*{1. Dependence of $\Gamma$ on $\Plamp$}
We observed empirically that $\GammaP$ increased approximately linearly with $\Plamp$, the power dissipated by the manganin resistor ``lamp''. 
The emission spectrum of the lamp depends on the temperature of the lamp, which is determined by a balance of the dissipated heat and the thermal conductivity to the nearest cold heat sink (likely the mixing chamber).
We expect dissipation to be uniform in the manganin filament of the lamp and independent of temperature for the range of our experiment.
On the other hand, the thermal conductivity of the leads which determines the temperature of the lamp filament is itself temperature-dependent and may vary along the length of the leads.

As described Section V A, the lamp was required to compare QP generation that occurs with parity switching due to $\PAT$ ($\gpat$) and QP generation that does not occur with parity switching ($\gother$). 
For the purpose of constraining $\gother$, measurement at any increased $\GammaP$ without increased device temperature was sufficient, since a second set of \{$\Gamma_P(\Phi), \Gamma_N(\Phi)\}$ constrained the model and fixed the strength of the trapping rate.

\begin{figure}[h]
\includegraphics[width=1.0\columnwidth]{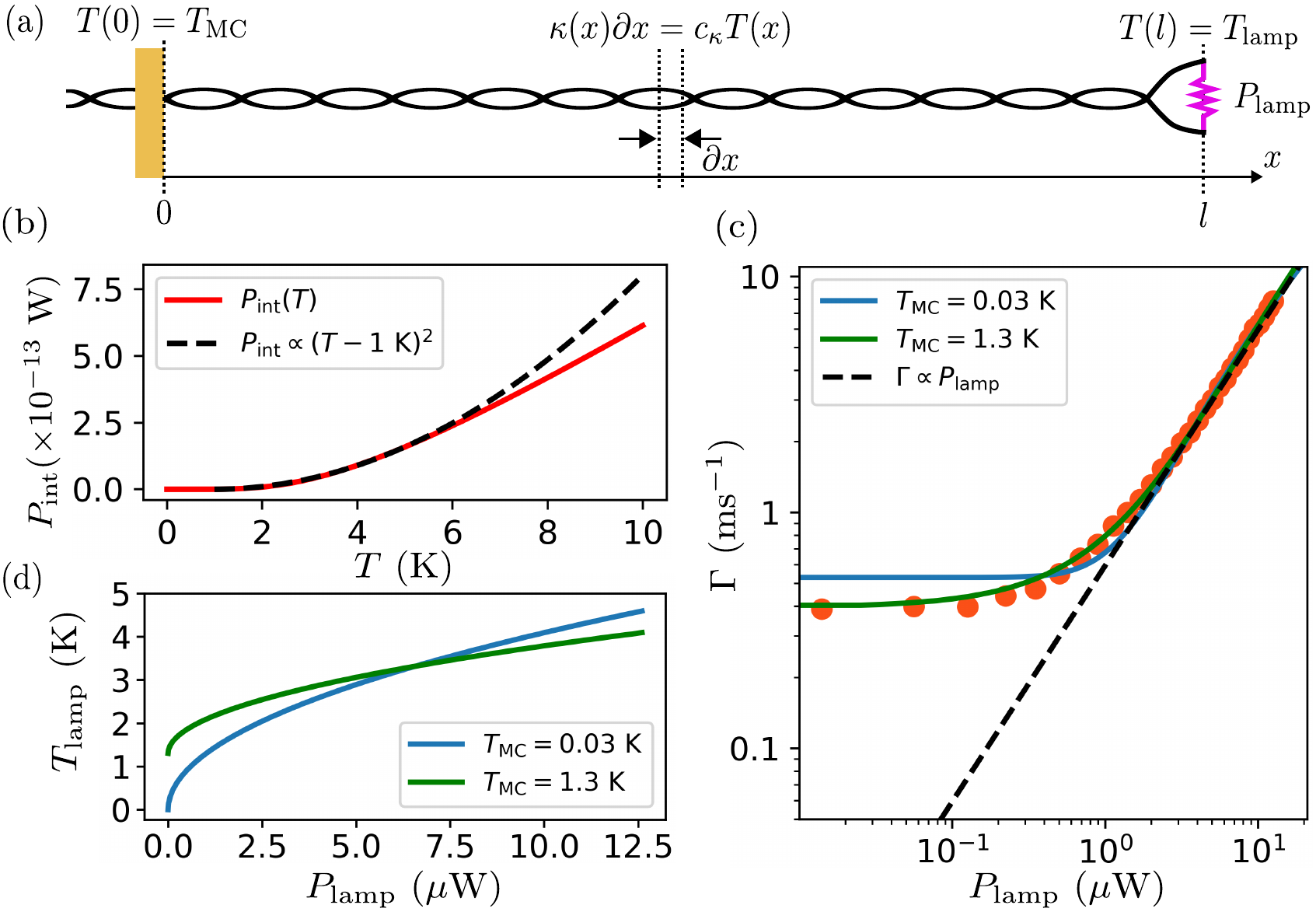} 
\caption{ \label{fig:lampfig}
(a) Diagram of model for the temperature of the lamp $\Tlamp$ as a function of the power dissipated by the lamp $\Plamp$.
The thermal conductivity per unit length of the leads is assumed to be linearly proportional to the temperature, which varies as a function of position $x$ along the leads. 
The temperature at the mixing chamber $T(0) = T_\mathrm{MC}$ is approximately 30 mK. 
(b) Power radiated by a 3D blackbody at temperature $T$ integrated from 100 to 300 GHz. 
We observe that from 1 K to 5 K, the dependence is approximately quadratic in temperature with an offset of 1 K. 
At higher temperatures, $P_\mathrm{int}\propto T$.
(c) Measured $\Gamma$ as a function of the power dissipated by the lamp (orange).
The data is the same as Fig. 4(b) but here shown in log-log scale. 
We fit it with the model in which $\Gamma \propto P_\mathrm{int}(\Tlamp)$, where $\Tlamp$ depends on $\Plamp$ according to Eq.~\eqref{eq:TlampPlamp}.
In blue, we fit assuming the lamp is at the mixing chamber temperature when no power is dissipated ($T_\mathrm{MC} = 0.03 \ \mathrm{K}$).
The model captures the linear behavior between $\approx 1 - 10 \ \mu \mathrm{W}$ but not the low-power dependence of $\Gamma$.
In green, in an attempt to fit $\Gamma(\Plamp<1 \ \mu \mathrm{W})$, we allow $\Tlamp$ at $\Plamp=0 \ \mu\mathrm{W}$ to vary as a fit parameter, which yields $T_\mathrm{MC} = 1.3 \ \mathrm{K}$. 
This is an unreasonably high temperature, indicating that this model does not accurately capture the low power behavior. 
(d) The temperature of the lamp $\Tlamp$ as a function of $\Plamp$, calculated using the fit parameters of (b).
In blue, $\Tlamp(0)$ is assumed to be the mixing chamber temperature, while in green, $\Tlamp(0) = 1.3 \ \mathrm{K}$.
In both cases, $\Tlamp \propto \Plamp^{1/2}$. 
}
\end{figure}

Nonetheless, we attempted to understand the linear dependence of $\GammaP$ on the power dissipated by the lamp $\Plamp$ with a simple model that takes into account temperature-dependence of the thermal conductivity of the NbTi/Cu leads. 
We assume that the heat flow in the wire is constant in the steady-state, but the thermal conductivity of the Cu in the leads depends on the position $x$ from the mixing chamber due to the temperature gradient [Fig.~\ref{fig:lampfig}(a)]
\begin{equation}
\kappa(x) \frac{\partial T}{\partial x} \partial x = \Plamp  .   
\end{equation}
The temperature of the lamp $\Tlamp$ is the temperature at $T(x = l)$, where $l$ is the length from the mixing chamber where the NbTi/Cu leads are thermalized to the lamp.
The thermal conductance per unit length $\kappa(x)$ depends on temperature according to the Weidemann-Franz law, $\kappa(x) \partial x = c_\kappa T(x)$. 
The proportionality factor $c_\kappa$ is related to the electrical resistivity $\rho$ and the Lorenz number $L$, $c_\kappa = L/\rho$, and we treat it as a fit parameter here. 
Substituting for $\kappa$ and integrating, we find that $\Tlamp \propto \Plamp^{1/2}$,
\begin{equation}
\Tlamp = \sqrt{\Plamp \frac{2l}{c_\kappa}+T_\mathrm{MC}^2}.\label{eq:TlampPlamp}
\end{equation}
The mixing chamber temperature is $T_\mathrm{MC} \approx 30 \ \mK$.

Next, we propose that $\GammaPAT$ is proportional to the integrated power $P_\mathrm{int}$ radiated from the lamp that enters the cavity and causes $\PAT$.
Treating the lamp as a 3D blackbody, we find that integrating the radiated power from $100-300 \ \mathrm{GHz}$ yields a dependence on the temperature of the lamp ($\Tlamp$) that is approximately quadratic for temperatures $\sim 1-5 \ \mathrm{K}$ [Fig.~\ref{fig:lampfig}(b)].
The upper bound of this frequency range of photons which induce $\PAT$ is determined by the constraint that the size of the qubit must be smaller than the wavelength of the photon. 
Higher frequency photons with smaller wavelengths are increasingly likely to be absorbed in the pads of the device due to the larger absorption area. 

In Fig.~\ref{fig:lampfig}(c), the measured $\GammaP$ is shown as a function of $\Plamp$.
We fit the data with a simple model in which the lamp radiates the spectral density of a 3D blackbody at $\Tlamp$, and $\GammaP$ is proportional to $P_\mathrm{int}$, the integrated power between 100 and 300 GHz radiated by the lamp:
$\GammaP = AP_\mathrm{int}(\Tlamp)+B$.
In this model $A$ is an unknown proportionality constant between $\GammaP$ and $P_\mathrm{int}$ which includes attenuation and coupling of radiation to the qubit which is assumed to be frequency-independent for simplicity. 
This scale factor also includes the additional $\Gamma$ resulting from increased $\GammaQPT$ due to the QPs generated by the increased $\GammaPAT$ (when $g_P \gg \gother$, $\GammaQPT $ is approximately proportional to $\GammaPAT$). 
The lamp-power-independent offset $B$ accounts for background $\PAT$ from other sources and $\QPT$ from QPs existing at $\Plamp = 0 \ \mu \mathrm{W}$.
Combining  $\Tlamp \propto \Plamp^{1/2}$ with $P_\mathrm{int} \propto \Tlamp^2$, we expect a linear dependence of $\GammaPAT$ on $\Plamp$ for a range corresponding to $\Tlamp \approx 1 - 5 \ \mathrm{K}$.
In the range of $\Plamp \approx 1 - 12.6\  \mu\mathrm{W}$, the data shows this linear dependence. 

The blue line shows a fit with $T_\mathrm{MC} = 0.03 \ \mathrm{K}$, which models the linear part of the data reasonably well but does not capture the low power behavior. 
The fit parameter $c_\kappa$ corresponds to an electrical resistivity of $\rho = 2.9 \times 10^{-9} \ \Omega \mathrm{m}$. 
Shown on log-log scale, it is clear that $\GammaP(\Plamp)$ is not linear for $\Plamp \lesssim 1 \ \mu \mathrm{W}$.
The fit to this range can be improved by allowing $T_\mathrm{MC}$ to vary (green), but yields $T_\mathrm{MC} = 1.3 \ \mathrm{K}$, which is  forty times hotter than the mixing chamber. 
Thus, while this model does capture the observed linear behavior for $\Plamp \approx$ 1 -- 12.6 $\mu \mathrm{W}$, the low $\Plamp$ behavior is not described by this simple model. 
This is reasonable given that the model assumed $\Tlamp \gg T_\mathrm{MC}$ and $g_p \gg \gother$.

\subsection*{2. Effect of lamp on temperature}
\begin{figure}[h]
\includegraphics[width=1.0\columnwidth]{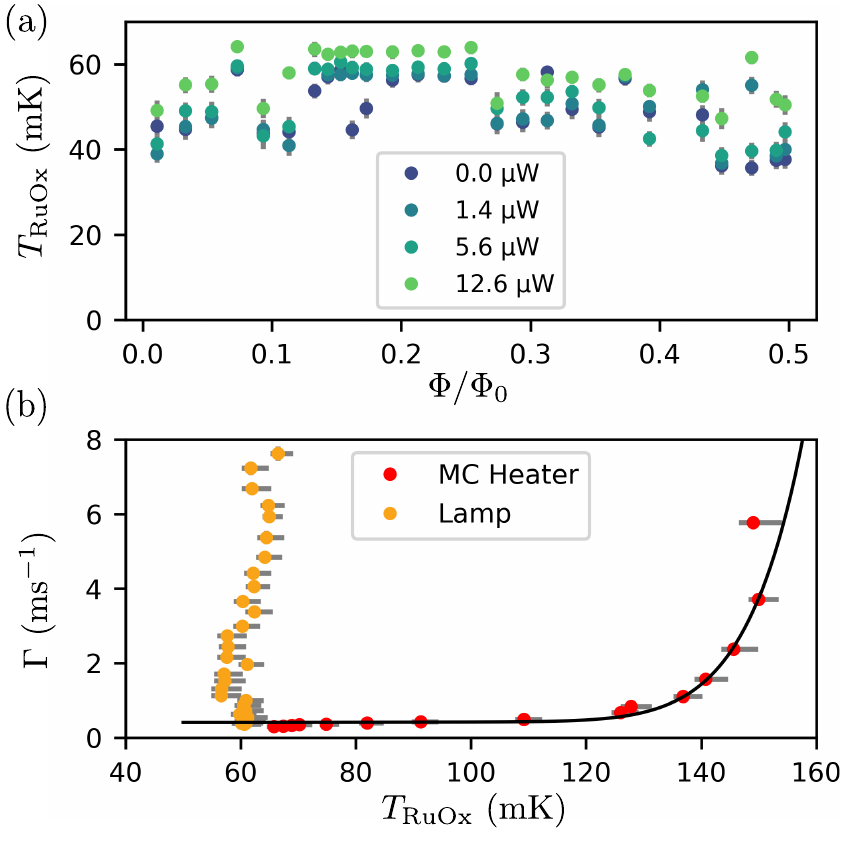} 
\caption{ \label{fig:tempfig}
(a) Temperature measured by RuOx thermometer for the data of Fig. 4(c). 
The colors correspond to different lamp powers. 
The thermometer is mounted on the copper bracket supporting the 3D readout cavity in which the qubit is embedded. 
The temperature remains well below the temperature at which thermal QPs are generated for all values of $\Phi/\Phi_0, \Plamp$. 
(b) The measured parity-switching rate $\Gamma$ as a function of temperature measured by the RuOx thermometer near the readout cavity. 
Two sweeps are shown.
In yellow, the lamp power is swept [Fig.~\ref{fig:tempfig}(b)], and here we observe the bracket temperature does not change significantly as $\Gamma$ increases by this mechanism.
In red, the mixing chamber temperature is swept by increasing power dissipated by heaters on the mixing chamber plate. 
The temperature measured by the RuOx increases as the mixing chamber temperature rises, but $\Gamma$ doesn't increase dramatically until approximately 125 mK. 
This shows that the slight temperature increases shown in (a) cannot be responsible for the large increase in $\Gamma$. 
The black line shows a fit to $\Gamma$ vs. $T$, from which we may extract the average gap $\bar{\Delta}/h\approx 51.8 \ \mathrm{GHz}$.
}
\end{figure}

To confirm that the manganin lamp was increasing $\GammaP$ by $\PAT$ rather than by increasing the temperature of the device and generating additional thermal QPs, we mounted a ruthenium oxide (RuOx) thermometer on the bracket holding the copper cavity. 
The temperature measured by this thermometer is plotted for each measurement of $\GammaP$ in Fig.~\ref{fig:tempfig}(a).
The difference between the temperature measured by this thermometer and the mixing chamber temperature $T_{\mathrm{MC}} \approx 30 \ \mathrm{mK}$ attributed to the thermal resistance from the end of the bracket to the mezzanine to the mixing chamber.  
We observe that while the temperature increases by several mK as the power dissipated by the lamp increases, this increase is far too small to generate a significant number of thermal QPs. 
This was verified by performing a separate sweep in which we increased the temperature by a heater mounted on the mixing chamber plate [Fig.~\ref{fig:tempfig}(b), red]. 
We can see that when the bracket temperature measured by the RuOx increased due to the remote heater by similarly small amounts, there was minimal change in $\GammaP$. 
Only when the bracket RuOx reached approximately 150 mK did the increase in $\GammaP$ match what was produced by $12.6 \ \muW$ dissipated by the lamp.  
Fig.~\ref{fig:tempfig}(b) shows that the increase in $\GammaP$ due to power dissipated by the lamp cannot be explained by an increased temperature of the sample. 

This dependence of $\Gamma$ on the mixing chamber temperature was used to estimate the average gap of the two superconducting films $\bar{\Delta}=(\Delta_L+\Delta_H)/2$, since thermal activation of QPs will be sensitive to this value.
The black solid line shows a fit to a model in which there is temperature-independent $\PAT$ and temperature-dependent $\QPT$ due to the QP energy distribution changing and thermal QPs being generated at higher temperature. 
The only fit parameters used in this model are $\GammaPAT$ and $\bar{\Delta}$.
The fitted $\bar{\Delta}$ is not sensitive to the assumed values for trapping and recombination rates that are fixed parameters in the model. 
In this case, we set $\gother = 0$, since the data cannot distinguish between a temperature-independent $\PAT$ rate vs. a temperature-independent background of excess QPs. 
Repeating the fit with $\gother$ as a fit parameter and $\GammaPAT = 0$ yielded the same $\bar{\Delta}$. 
In each case, we found $\bar{\Delta}/h = 51.8 \ \mathrm{GHz}$, which is consistent with reported superconducting gap measurements of thin-film aluminum~\cite{chubov_dependence_1969,court_energy_2007}.

\section*{appendix g: Fit sensitivity to trapping rate $s$}
\begin{figure}[h]  
\includegraphics[width=1.0\columnwidth]{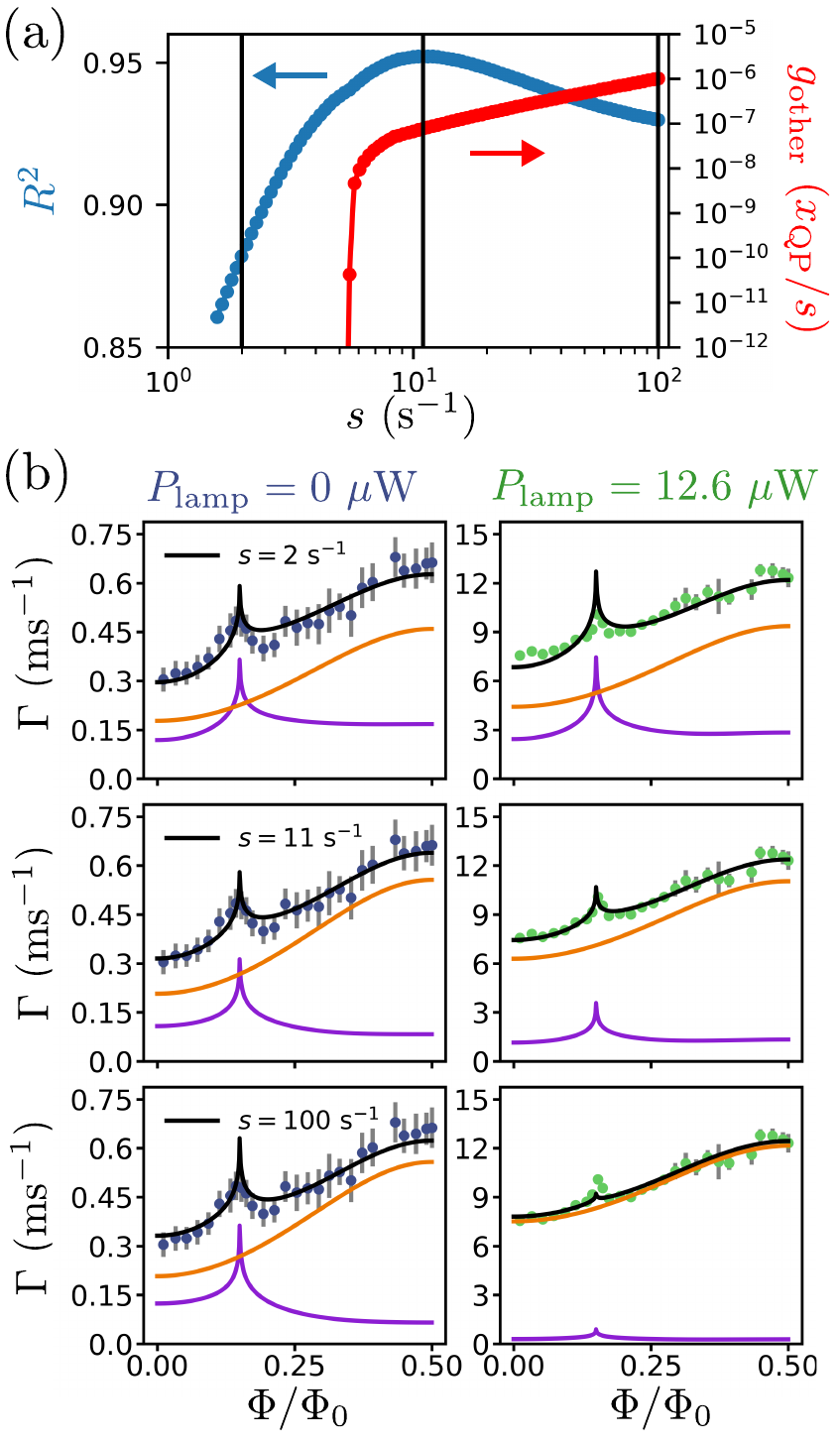} 
\caption{ \label{fig:fitparam}
(a) Pseudo-$R^2$ goodness-of-fit metric for fits to the full data set as a function of trapping rate $s$ (blue, left axis).
The value of $\gother$ obtained in each fit is also plotted for reference (red, right axis). 
Vertical black lines indicate values of $s$ for which the fits are shown in (b). 
(b) Fits to the $\Plamp = 0 \ \mathrm{\muW}$ (left) and $\Plamp = 12.6 \ \mathrm{\muW}$ (right) data for the values of $s$ marked by vertical lines in (a). 
The fits corresponding to lower values of $R^2$ (i.e. $s=2\ \mathrm{s^{-1}}, s = 100 \ \mathrm{s^{-1}}$) appear to describe the intermediate peak at $\Phi/\Phi_0 = 0.145$ less well for the $\Plamp = 12.6 \ \mathrm{\muW}$ as compared to the best fit ($s=11 \ \mathrm{s^{-1}})$.
 }
\end{figure}
Fig.~\ref{fig:fitparam} demonstrates the sensitivity of the fit shown in Fig.~4(c) to the trapping rate $s$.
In Fig.~\ref{fig:fitparam}(a), we plot the pseudo-$R^2$ goodness-of-fit metric for fits to the full data set fixing $s$ (blue, left axis).
We use the definition $R^2 = \sum_i [1-(S_{res,i}/S_{tot,i})]/4 , \ i \in \{0,1,2,3\}$, where $S_{res,i}$ and $S_{tot,i}$ are the sum of squares difference between the data and the model ($S_{res,i}$) and between the data and mean ($S_{tot,i}$) measured at the $i$th lamp power.

We observe a maximum of $R^2$ near $s = 11 \ \mathrm{s^{-1}}$ corresponding to the best fit.
For $s\lesssim 6 \ \mathrm{s^{-1}}$, generation by $\PAT$ is sufficient to generate the approximate $\xqp$ in the device, such that the contributions of $\gother$ found by the fits in this range are negligible.
For  $s\gtrsim 6 \ \mathrm{s^{-1}}$, the best fit $\gother$ increases linearly with $s$ in order to keep the $\xqp$ at the value which best fits the data for $\Plamp = 0 \ \mathrm{\mu W}$.

Fig.~\ref{fig:fitparam}(b) illustrates how values of $s$ differing from the best fit value affect the model. 
We observe that for a lower value of $s=2 \ \mathrm{s^{-1}}$, the additional QPs generated by the enhanced $\GammaPAT$ with $\Plamp = 12.6 \ \mathrm{\mu W}$ cause a level of $\GammaQPT$ that gives a peak at $\peakfluxB$ which is too large compared to the data.
For a higher value of $s=100 \ \mathrm{s^{-1}}$, the strong trapping rate suppresses the additional QPs generated by the enhanced $\GammaPAT$ with $\Plamp = 12.6 \ \mathrm{\mu W}$, causing a level of $\GammaQPT$ that gives a peak at $\peakfluxB$ which is too small compared to the data.

\section*{appendix h: direct observation of quasiparticle bursts}
One possible source of pair-breaking energy for QP generation (i.e., a contribution to $\gother$) is ionizing radiation.
The energy cascade from the ionization of atoms in the superconducting films or substrate, described in~\cite{vepsalainen_impact_2020,martinis_saving_2021, wilen_correlated_2021, mcewen_resolving_2022}, ultimately results in ``bursts'' of QPs. 
Experimentally, evidence for these bursts has thus far come in the form of sudden drops in $T_1$ in superconducting qubits or drops in the quality factor of superconducting resonators.
In this device, we found evidence of rapid parity switching which further substantiates the hypothesis that these sudden drops in $T_1$ are due to QPs. 
By tuning the flux of our SQUID offset-charge-sensitive transmon such that $E_J/E_C \lesssim 20$, we were able to measure the parity directly via the parity-dependent dispersive shift as described in Ref.~\cite{serniak_direct_2019}. 
With a quantum-limited SNAIL parametric amplifier~\cite{frattini_optimizing_2018}, we were able to discern the parity with a single $4 \ \mu\mathrm{s}$ measurement of the readout cavity. 

\begin{figure*}[hbt!]
\includegraphics[width=2.0\columnwidth]{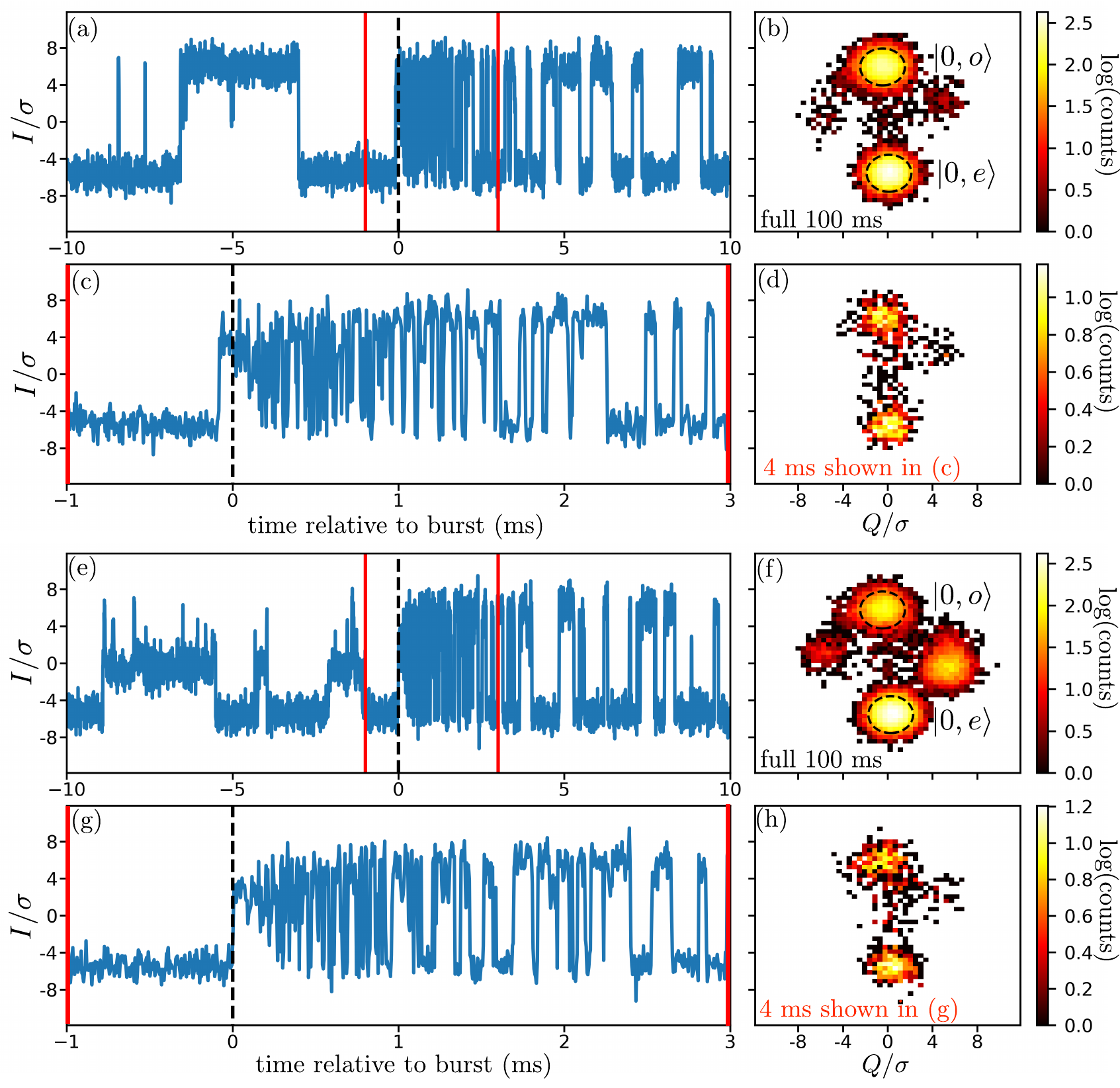} 
\caption{ \label{fig:burstfig}
(a) Time series of the $I$ quadrature of qubit-state measurements repeated every $5 \ \mathrm{\mu s}$, in units of the $\sigma$ of the Gaussian distributions of the ground-state measurement histograms. 
Rapid switching between parity states begins at $t=0$, which we attribute to a ``burst'' of QPs generated by ionizing radiation.
(b) 2D histogram of qubit-state measurements in the complex plane for the full 100 s time series from which the data in (a) is drawn, shown for state identification.
Most of the counts form two Gaussian distributions with positive $Q$, which we attribute to the even and odd parity ground states, respectively.  
The counts at negative $Q$ correspond to excited states of the qubit. 
Data is plotted in log scale for visibility of excited states, and dashed lines indicate $2 \sigma$. 
(c) Zoom in on (a) showing 4 ms around the onset of the burst.
The value jumps to a value between $|0,e\rangle$ and $|0,o\rangle$ due to switching that is fast compared to the $1/(5 \ \mathrm{\mu s})$ measurement repetition rate. 
This is followed by rapid switching between the parity states which decays in frequency over time.
(d) Same as (b), but for only the data in the window between the red lines shown in (c). 
The data is primarily in the ground states, showing that the data near $I\approx 0$ at $t = 0$ is due to rapid switching rather than a jump to an excited state.
(e-f) are analogous to (a-d), but show a burst correlated with a jump in $n_g$ due to the redistribution of charge in the substrate resulting from the ionizing impact.
At $t<0$, the $|0,o\rangle$ state is located at $I/\sigma \approx 0$, $Q/\sigma \approx 6$, and parity switches occur between the even and odd parity ground states in this configuration.
When the burst occurs at $t = 0$, the $|0,o\rangle$ state jumps to the same configuration as in (a-d), indicating a jump in $n_g$ correlated with a burst.
}
\end{figure*}

In addition to generating high-energy phonons which break Cooper pairs and generate QPs, ionizing radiation impacts can also cause $n_g$ jumps due to the redistribution of charge in the substrate. 
With this direct detection technique, which tracks parity switching and charge jumps simultaneously, we were able to quantify how many of the observed burst events are correlated with an $n_g$ jump. 
In Fig.~\ref{fig:burstfig}, we show two examples of bursts: one in which the $n_g$ is unchanged (a-d) and one in which $n_g$ changes visibly before and after the burst (e-h).

In Fig.~\ref{fig:burstfig}(a), we show a 20 ms-long jump trace of qubit measurements, with the $I$ quadrature of the complex-valued signal plotted. 
Fig.~\ref{fig:burstfig}(b) shows the 2D histogram in the complex plane of the full 100 ms time series from which the data in (a) is drawn, which consists of $2\times 10^5$ qubit measurements. 
The readout pulse length and integration time were $4 \ \mu \mathrm{s}$, with $1 \ \mu \mathrm{s}$ delay between measurements. 
We can see the measurements primarily form two Gaussian distributions near $Q$ = 0, which we interpret as the even and odd parity ground states (even and odd are assigned arbitrarily).
The remaining population corresponds to higher energy states.
At $t < 0$, we can see from the value of $I$ that the qubit initially has even parity and then switches to odd parity at $t\approx -7 \ \mathrm{ms}$ and back to even at $t = -3 \ \mathrm{ms}$. 
The background parity-switching rate at this flux value was $\Gamma \approx 600 \ \mathrm{s^{-1}} \approx 1/1.7 \ \mathrm{ms}$ during this cooldown.

Fig.~\ref{fig:burstfig}(c) zooms in on a 4 ms section around a burst event, which we label as occurring at $t=0$. 
We observe that at $t = 0$, the value of $I$ jumps to a value between the values corresponding to $|0,e\rangle$ and $|0,o\rangle$. 
This is likely due to the parity switching much faster than the $5 \ \mu \mathrm{s}$ repetition time of our measurement. 
We confirm that this is not a result of qubit excitation by checking the 2D histogram of points during this 4 ms segment, in which it is clear that smearing between $|0,e\rangle$ and $|0,o\rangle$ is much more prevalent than qubit excitation [Fig.~\ref{fig:burstfig}(d)]. 
Following this jump, the qubit state switches between the $|0,e\rangle$ and $|0,o\rangle$ states much faster than the  $1.7 \ \mathrm{ms}$ lifetime characteristic of the majority of the jump trace. 
The parity-switching rate appears to decay significantly by $t=10 \ \mathrm{ms}$, although it is not clear that it has returned to the consistent background rate.

Fig.~\ref{fig:burstfig}(e-h) are analogous to (a-d), except that at $t=0$, there is a sudden change in the dispersive shifts of the qubit states signalling a jump in $n_g$ in addition to the onset of rapid parity-switching.   
In this data set, we search for bursts when the configuration of $n_g$-dependent dispersive shifts provide maximal distinguishability of the even and odd parity states (the configuration of (a-d)).
In the event shown in Fig.~\ref{fig:burstfig}(e-h), the dispersive shift of the $|0,o\rangle$ state changes at $t=0$, changing from $I/\sigma \approx 0$ to the optimal configuration in which $I/\sigma \approx 6$. 
Impacts which cause $n_g$ jumps out of this optimal configuration into others will not be detected; however, we note that they will occur at the same rate as jumps into the optimal configuration. 

We observed approximately 209 events in ${\approx 5.6}$  hours of data. 
Of the 209 impacts, we find that 60 are correlated with an $n_g$ jump, which provides additional information about the location of the impacts.
Adding in an equal number of undetected jumps out of the optimal configuration, we estimate an event rate of $\approx 1/75 \ \mathrm{s}$.
This is consistent the impact rates reported in Refs.~\cite{wilen_correlated_2021,grunhaupt_loss_2018, mcewen_resolving_2022} after taking into account the area of the substrate (our sapphire substrate is 3 mm x 15 mm). 
The extent to which bursts contribute to the background excess QP density and parity-switching rate depends on the energy deposited by the ionizing radiation, the frequency of the events, and the timescale on which the density decays.
Further work is necessary to estimate the energy deposited by these events and to understand the decay of the QP density after a burst, which may involve several timescales~\cite{martinis_saving_2021,mcewen_resolving_2022}.

\bibliography{references.bib}
\end{document}